\documentclass[12pt]{iopart}
\usepackage{epsf}
\usepackage{epsfig}
\usepackage{amssymb}

\begin{document}

\title[Out-of-equilibrium Edwards-Wilkinson equation]
{Out-of-equilibrium relaxation of the Edwards-Wilkinson elastic line}
\author{Sebastian Bustingorry$^1$, Leticia F. Cugliandolo$^2$
and Jos\'e Luis Iguain$^3$}
\address{$^1$DPMC-MaNEP, Universit{\'e} de Gen{\`e}ve,
24 Quai Ernest Ansermet, 1211 Gen{\`e}ve 4, Switzerland\\
$^2$Universit\'e Pierre et Marie Curie -- Paris VI, LPTHE UMR 7589,
4 Place Jussieu, 75252 Paris Cedex 05, France\\
Departamento de F\'{\i}sica, FCEyN, Universidad Nacional de Mar del Plata\\
De\'an Funes 3350, 7600 Mar del Plata, Argentina}
\date{\today}

\begin{abstract}
We study the non-equilibrium relaxation of an elastic line described
by the Edwards-Wilkinson equation. Although this model is the
simplest representation of interface dynamics, we highlight that many
(not though all) important aspects of the non-equilibrium relaxation
of elastic manifolds are already present in such quadratic and clean
systems. We analyze in detail the aging behaviour of several two-times
averaged and fluctuating observables taking into account finite-size
effects and the crossover to the stationary and equilibrium
regimes. We start by investigating the structure factor and
extracting from its decay a growing correlation length. We present
the full two-times and size dependence of the interface roughness and
we generalize the Family-Vicsek scaling form to non-equilibrium
situations. We compute the incoherent scattering function and we
compare it to the one measured in other glassy systems. We analyse
the response functions, the violation of the fluctuation-dissipation
theorem in the aging regime, and its crossover to the equilibrium
relation in the stationary regime. Finally, we study the
out-of-equilibrium fluctuations of the previously studied two-times
functions and we characterize the scaling properties of their
probability distribution functions.  Our results allow us to obtain
new insights into other glassy problems such as the aging behavior in
colloidal glasses and vortex glasses.

\end{abstract}



\maketitle


\section{Introduction}
\label{sec:intro}

\subsection{Aim}

Interfaces are important in many physical, chemical and biological
phenomena.  In the physical context they appear in stochastic surface
growth~\cite{Barabasi-Stanley}, domain growth and coarsening phenomena
(as domain walls)~\cite{Alan}, type-II superconductivity (as magnetic
flux lines)~\cite{rusos}, fracture cracks~\cite{cracks}, and fluid
invasion in porous media~\cite{poros}, among others
realizations. Although specific problems involve different levels of
complexity, as the inclusion of non-linear or disorder contributions,
the main features of interface dynamics are already contained in the
simplest theoretical formulation of the problem: the Edwards-Wilkinson
(EW) equation~\cite{EW}.

The aim of this article is to exhibit, in a concrete and fully
solvable example, several generic properties of the averaged and
fluctuating aging dynamics of finite and infinite elastic
manifolds. The problem we analyze is the relaxation of an elastic line
described by the Edwards-Wilkinson equation~\cite{EW}. The embedding
space dimension plays an important role. In the particular one
dimensional transverse space case on which we focus here the dynamics
has aspects of diffusion, glassiness and saturation depending on the
time and length scales observed.

The relaxation process we are interested in is the
following~\cite{Leto}.  After equilibration at a temperature $T_0$ the
system is suddenly taken to a different working temperature $T$ that
can be higher or lower than $T_0$. Time is then set to zero. The line
subsequently tries to adapt to the working temperature. The relaxation
is characterized by the time-dependence of correlation and linear
response functions.  One lets the line relax until a waiting-time,
$t_w$, when the quantities of interest are first recorded and later
compared to their values at a subsequent time $t$.

A series of glassy properties of elastic lines evolving in disordered
environments have been recently reported in different contexts,
related to directed polymers in random
media~\cite{Yosh98,Bustingorryetal}, the vortex glass dynamics in high
temperature superconductors~\cite{us,schehr05,andrei}, and domain wall motion
in magnetic systems~\cite{Kolton,Spaniards}. All these studies focused
on models including disorder, which non-disordered counterparts belong
to the EW universality class of interface
dynamics~\cite{Barabasi-Stanley}, except for Ref.~\cite{Spaniards}
in which the analysis is focused on the non-linear contribution in the
Kardar-Parisi-Zhang (KPZ)~\cite{KPZ} universality class.  In order to
disentangle the effects of elasticity and disorder it is then
important to study in detail the dynamics of the non-disordered
counterparts.

In general, the glassy phenomenon in finite dimensional models of elastic lines
with and without quenched disorder appears as a dynamic
crossover~\cite{Yosh98,Bustingorryetal,us}.  For all waiting-times, $t_w$, that
are longer than a size, $L$, and eventually also temperature, $T$,
dependent {\it equilibration time}, $t_L$, the dynamics is
stationary. Instead, for $t_w < t_L$ the system is in the preasymptotic regime and the relaxation occurs out of
equilibrium as demonstrated by two-times correlations and linear
responses that age.
For each waiting-time the dependence on the time-delay, $\Delta
t\equiv t-t_w$, then shows a growth regime for $\Delta t < t_L$ and
saturation regime at longer time, $\Delta t>t_L$, where the correlation functions saturate to a size dependent value.

\begin{figure}[!tbp]
\centerline{
\includegraphics[angle=-0,width=10cm,clip=true]{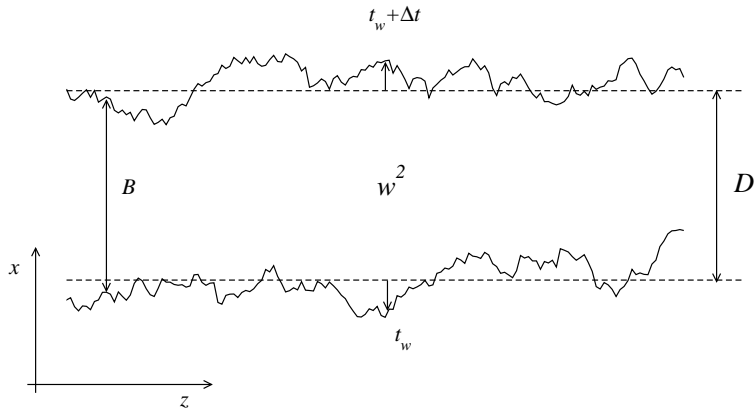}
}
\caption{\label{f:sketch_lines} Two line configurations at
different times, {\it e.g.} $x(z,t_w)$ and $x(z,t_w+\Delta t)$, showing the
relevant displacements defining the mean-squared-displacement of the
line segments, $B$, the roughness, $w^2$, and the
mean-squared-displacement of the center of mass, $D$.}
\label{fig:sketch-lines}
\end{figure}

Here we compute, analytically, the averaged two-times roughness and we
relate it to the displacement field, which was the main focus of
previous studies of the dynamics of elastic manifolds in random
environments~\cite{Yosh98,us,Cule}, and the center of mass evolution
of the interface (see the sketch in Fig.~\ref{fig:sketch-lines}).  The
quadratic character of the model allows us to control the crossover to
equilibration and saturation of finite lines.  We next deduce several
other correlators that are of interest in glassy dynamics. In
particular, we analyze the incoherent scattering function and compare
it to light scattering measurements in clay colloidal
suspensions~\cite{laponite}. We interpret our results in terms of a
two-times correlation length that we evaluate and confront to the one
measured in other aging glassy systems~\cite{Parisi,Ludovic}.  We also
study linear responses and their relation to the companion
correlations.  We discuss in detail the special features of these
two-times observables linked to the multiplicative -- diffusive --
scaling form.

Finally, we focus on thermally induced fluctuations.  Recently, the
importance of studying fluctuations -- and not only averaged
quantities -- in dynamic phenomena was stressed in several
contexts. R\'acz proposed to use scaling functions characterizing the
fluctuations of global observables in {\it non-equilibrium steady
states} to classify systems in `universality classes' dictated by
symmetries and dynamic mechanisms~\cite{Racz}. In {\it aging glassy}
systems the study of fluctuations seems to be fundamental to
understand the mechanism for the dramatic slowing down and
non-equilibrium relaxation. Chamon {\it et
al}~\cite{Chamon-etal,Chamon-Cugliandolo} proposed a symmetry based
sigma-model like theory for fluctuations in conventional glassy
systems. For a number of reasons this theory is not expected to apply,
without modification, to interface dynamics. In more technical terms,
the averaged interface dynamics is
characterized by a multiplicative aging scaling that should result in
the need to modify the approach in \cite{Chamon-etal} to take this
feature into account.  We thus wish to confront the fluctuations of
conventional glassy systems to those of interface models searching for
similarities and differences.  With this purpose we derive the
probability density functions (pdfs) of several two-times correlation
functions. We compare our results to previous studies~\cite{Zoltan},
which focused on the time-delay dependence of the 
fluctuations~\footnote{We note that the solution presented in~\cite{Zoltan}
corresponds actually to the $t_w=0$ solution in this article.},
and to predictions of the time-reparametrization invariance scenario
of glassy systems~\cite{Chamon-etal,Chamon-Cugliandolo}.

The problem addressed here is related to a number of other models that
have already been studied in the literature; among the pure cases one
has the Langevin dynamics of the Gaussian and massless scalar
field~\cite{Cukupa}, the XY ferromagnet~\cite{Cukupa,Behose}, the
$p=2$ spherical spin-glass~\cite{Cude} and the $O(N)$ ferromagnet in
the large $N$ limit~\cite{Chcuyo}. Related models with quenched
disorder are problems of elastic manifolds in quenched random
environments~\cite{Yosh98,Bustingorryetal,Kolton,Cule}.  In all these
studies the finite size dependence was not taken into account and the
dynamic fluctuations were not studied. We explain in the conclusions
how our results relate to the ones in these papers.

\subsection{The model}
\label{sec:model}

The EW equation for a scalar field $x$ representing the
height of a surface over a one dimensional substrate parametrized
by the coordinate $z$ (a one dimensional directed interface) is
\begin{eqnarray}
\label{eq:EW}
\partial_t x(z,t) = \nu \partial^2_z x(z,t)+f(z,t)+\xi(z,t), \\
\langle \xi(z,t) \rangle = 0, \qquad \langle \xi(z,t) \xi(z',t') \rangle =
\frac{2T}{\gamma} \delta(z-z') \delta(t-t'),
\end{eqnarray}
with $\nu=c/\gamma$, $c$ the elastic constant, $\gamma$ the friction
coefficient, $T$ the temperature of the thermal bath and $\langle
\cdots \rangle$ the average over the white noise $\xi$, {\it i.e.} the thermal average. The term
$f=h/\gamma$ represents the effect of a perturbing field that
couples linearly and locally to the height, $-h(z,t) x(z,t)$. One
can also consider other types of perturbations that couple to more
complicated functions of the height, as discussed in Sect.~\ref{sec:response}.

We are interested in characterizing the dynamics of elastic
lines with {\it finite} and {\it infinite} length $L$. Thus for
convenience we shall take periodic boundary conditions in the $z$-direction.
Following Antal and R\'acz we
introduce a Fourier representation of the position dependent
distance of the line from its average~\cite{Zoltan}
\begin{equation}
\label{eq:defFouriermodes}
\delta x(z,t) \equiv
x(z,t)-\overline{x}(t)=\sum_{n=-\infty}^{\infty} c_n(t) \; e^{i k_n z},
\end{equation}
with $k_n=2 \pi n/L$.
The overline indicates an average over the full line's configuration;
\begin{equation}
\overline{x}(t)=\frac{1}{L}\int_0^L dz \; x(z,t)
\end{equation}
being then the mean height -- or center of mass -- of the interface.
The Fourier coefficients are given by
\begin{equation}
c_n(t) = \frac{1}{L} \int_0^L dz \; \left[ x(z,t)-\overline{x}(t)
\right] \; e^{-i k_n z}
\end{equation}
and their evolution  is given by
\begin{eqnarray}
\label{eq:EWn}
\partial_t c_n(t) = - \nu k_n^2 c_n(t)+f_n(t)+\xi_n(t), \\
\langle \xi_n(t) \rangle = 0, \qquad \langle \xi_n(t) \xi_{n'}(t') \rangle =
\frac{2T}{\gamma L} \delta_{n,-n'} \delta(t-t')
\end{eqnarray}
for all $n \neq 0$
while $c_0(t)=0$ at all $t$.
Note that equation~(\ref{eq:EWn}) effectively describes the
position of a particle in a harmonic potential with spring constant
$\nu k_n^2$ which implies an
elastic constant softening with decreasing $n$ or increasing system
size $L$. The solution to equation~(\ref{eq:EWn}) is
\begin{equation}
c_n(t)=c_n(0)\; e^{-\nu k_n^2 t}+ e^{-\nu k_n^2 t}\int_0^t  dt' \;
e^{\nu k_n^2 t'} \, \left[ \xi_n (t')+f_n(t') \right].
\end{equation}
where we set the initial time to $t=0$.  The coefficients $c_n(0)$
encode the structure of the initial conditions.  We are interested in
the evolution after an instantaneous quench
from equilibrium at a generic temperature $T_0$ to the working temperature
$T$. We are then considering that the initial conditions satisfy
\begin{equation}
\langle c_n(0) c_m(0) \rangle_{ic}=|c_n(0)|^2 \, \delta_{n,-m},
\end{equation}
where $|c_n(0)|^2$ should reflect the equilibrium at $T_0$ (see below).
When the initial temperature is higher than the working
temperature, $T_0>T$, the initial state is more disordered than the
equilibrium one at $T$ and the line tends to `order' as time
elapses. On the other hand, when $T_0 < T$, temperature fluctuations
roughen the initial -- more ordered -- configuration of the line. A
special case is $T_0 = 0$, corresponding to a perfectly flat initial
configuration.

The EW energy is just the one of a massless scalar field
and the model does not have a finite temperature static
phase transition.

Through out the article we shall present some figures which 
highlight the main analytical results.  Without loss of
generality we set $\gamma=\nu=1$ in all the figures.  We distinguish
noise averaged from fluctuating quantities by enclosing the former
with angular brackets.  In our expressions we do not write explicitly
the $T$, $T_0$ and $L$ dependencies but one has to keep in mind that
they are, in principle, always present.

\subsection{Organization of the paper}

The organization of the paper is the following. In
Sect.~\ref{sec:model} we had introduced the model and our
conventions.  Section~\ref{sec:structfactor} deals with the two-times
structure factor and Sect.~\ref{sec:ell} with the associated two-times
correlation length. In Sect.~\ref{sec:roughness} we analyze the line's
roughness and in Sect.~\ref{sec:MSD} we study the displacement field as well as the
motion of the center of mass. In Sect.~\ref{sec:incoherent} we derive a
wave-vector dependent correlation inspired in the incoherent scattering function
typical of particles in interaction.
Section~\ref{sec:response} analyzes
different linear response functions and their relation to the
associated correlations that is to say the modifications of the
fluctuation-dissipation theorem (FDT). In Sect.~\ref{sec:fluctuations} we study thermally induced
fluctuations by computing the probability distribution function
of the roughness, displacement and linear responses and we relate
to previous studies of elastic lines in equilibrium as well as
with the proposal for fluctuations in aging systems
based on the development of time-reparametrization invariance.
Finally, in Sect.~\ref{sec:summary} we summarize our results and
we present our conclusions.
The main features of the chosen observables are shown
in figures that display the {\it analytic} solution.

\section{Two-times structure factor}
\label{sec:structfactor}

A key quantity in the two-times evolution of the elastic line
is the local displacement of the surface height, defined as
$u(z,t,t_w)\equiv x(z,t)-x(z,t_w)$, which allows for the two-times generalization of several quantities.
The structure factor associated to this displacement is
\begin{eqnarray}
\label{eq:Sdeu}
& \langle S_n \rangle (t,t_w) =L \left\langle |c_n(t)-c_n(t_w)|^2
\right\rangle 
\nonumber\\
 & \quad = \frac{T_0}{\gamma \nu k_n^2} \left(1- e^{- \nu k_n^2 |\Delta t|}
\right)^2 \, e^{-2 \nu k_n^2 t_w}
\nonumber \\
 &
\quad +
\frac{T}{\gamma \nu k_n^2} \left[
2\left( 1-e^{-\nu k_n^2 |\Delta t|} \right)
\left( 1-e^{-2 \nu k_n^2 t_w} \right)
+\left( 1-e^{-2 \nu k_n^2 |\Delta t|} \right) e^{-2 \nu k_n^2 t_w} \right].
\nonumber \\
\end{eqnarray}
The structure factor defined in this way is the two-times
generalization of the one commonly used in the solution of the EW
equation~\cite{Kolton,loschinos}; by definition, it is symmetric
under $t \leftrightarrow t_w$. Hereafter we take $t\geq t_w$, dropping
the absolute value in the exponentials.

One can easily obtain several limits that show the
complexity of the aging behavior.
Let first consider the stationary limit, which is reached when
$\nu k_n^2 t_w\gg 1$ for all $n$ in (\ref{eq:Sdeu}). This condition is fulfilled when $t_w \gg t_L$, where
\begin{equation}
t_L\equiv \frac{L^2}{4\pi^2\nu}
\end{equation}
is a characteristic time marking the onset of finite size equilibration, and which will play an important role in the expressions below. Then, in the stationary limit, the structure factor becomes
\begin{equation}
\lim_{\nu k_n^2 t_w\gg 1}
\langle S_n \rangle( t,t_w) = \frac{2 T}{\gamma \nu
k_n^2} \left( 1-e^{-\nu k_n^2 \Delta t} \right).
\end{equation}

If one subsequently takes the small wave-vector limit for a given $\Delta t$ , {\it i.e.} $\nu k_n^2
\Delta t \ll~1$, the structure factor reaches the generic
asymptote $\langle S_n \rangle = 2T\Delta t/\gamma$. On the other hand, if one takes
the large wave-vector limit for a given $\Delta t$, {\it i.e.} $\nu k_n^2 \Delta t \gg 1$, one finds
the power-law decay of the structure factor
\begin{equation}
\lim_{\stackrel{\nu k_n^2 t_w\gg 1}{\nu k_n^2 \Delta t \gg 1}}
\langle S_n \rangle ( t, t_w) =
\frac{2T}{\gamma\nu k_n^2},
\label{eq:largek}
\end{equation}
typically found in harmonic processes.  Thus, this is the typical
stationary solution of the structure factor, displaying the $\Delta
t$-dependent saturation in the small wave-vector limit and the $1/k_n^2$
behavior in the long wave-vector limit corresponding to the modes
already equilibrated at the temperature $T$ \cite{Kolton}. When
$\Delta t \gg t_L$ all the modes are equilibrated at the working
temperature and the structure factor shows the power-law behavior in
(\ref{eq:largek}) for all $k_n$, without the saturation
regime. This behaviour indicates that the
large wave-vectors equilibrate first at the working temperature, and
then, for increasing time-delay, the number of equilibrated modes
increase. This is obviously related to a growing correlation length,
which will be analyzed in Sec.~\ref{sec:ell}, together with its aging
behaviour.

Now, let see how the waiting-time dependence
modifies the behaviour of the structure factor. First one notices that
the small wave-vector saturation asymptote does not depend on the
waiting time, {\it i.e.} the saturation value
\begin{equation}
\lim_{{\nu k_n^2 \Delta t \ll 1}}
\langle S_n \rangle(t, t_w) = \frac{2T}{\gamma} \Delta t,
\label{eq:smallk}
\end{equation}
is $t_w$-independent but time-delay-dependent.
On the other hand, the waiting-time is most prominent in the
asymptotic power-law regime, $\nu k_n^2 \Delta t \gg 1$, where one finds
\begin{equation}
\lim_{\nu k_n^2 \Delta t\gg 1} \langle S_n \rangle( t,t_w)
=
\frac{T_0}{\gamma \nu k_n^2} e^{-2 \nu k_n^2 t_w}+ \frac{T}{\gamma \nu
k_n^2} \left( 2-e^{-2 \nu k_n^2 t_w} \right).
\label{eq:waiting-time-dep-largek}
\end{equation}
From this expression one observes that the asymptotic power-law regime
in the large wave-vector limit changes from the $(T_0+T)/(\gamma \nu
k_n^2)$ asymptote at small waiting-time, to the $2 T/(\gamma \nu
k_n^2)$ asymptote in the long waiting-time limit.  For a fixed
time-delay value such that $\Delta t > t_w$, and for modes such that
$\nu k_n^2 \Delta t \gg 1$, the power-law behaviour changes from
$(T_0+T)/(\gamma \nu k_n^2)$ at lower values of $k_n$ to $2 T/(\gamma
\nu k_n^2)$ at higher values, indicating that the large wave-vectors
$\nu k_n^2 t_w \gg 1$ are equilibrated at the working temperature,
while the other modes still reflect the initial condition.  This also
implies the existence of a waiting-time-dependent mode, $k_w \sim (\nu
t_w)^{-1/2}$, which separates these two regimes, allowing us to write the
structure factor, in the asymptotic time-delay limit, in a scaling
form:
\begin{eqnarray}
& &
\lim_{\nu k_n^2 \Delta t\gg 1}
\langle S_n\rangle(t,t_w)
=
k_w^{-2} \; {\cal S}(k/k_w).
\label{eq:scaling-S}
\end{eqnarray}

\begin{figure}[!tbp]
\centerline{
\includegraphics[angle=-0,width=6cm,clip=true]{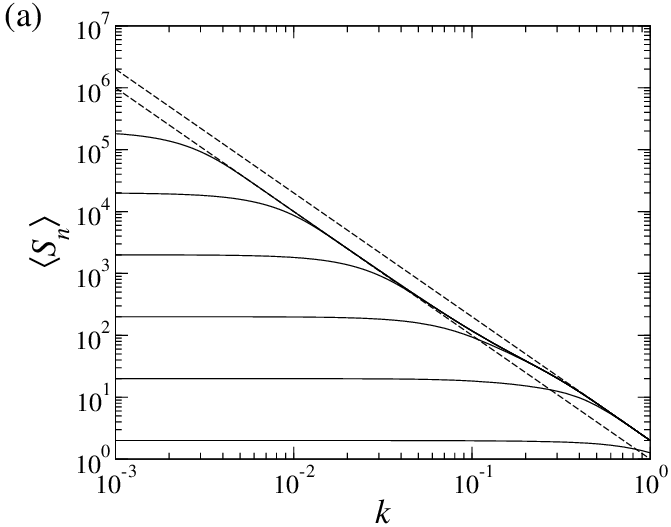}
\hspace{0.25in}
\includegraphics[angle=-0,width=6cm,clip=true]{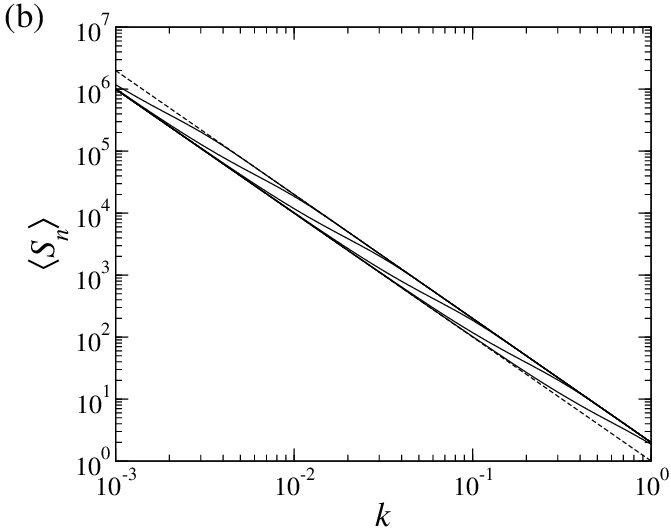}
}
\caption{\label{f:Sun_T00} $T=1$ evolution of the structure factor
from a flat initial condition ($T_0=0<T$).  (a) $t_w = 10$ and the
curves correspond to the time-delay values $\Delta t =
1,\,10,\,10^2,\,10^3,\,10^4$, and $10^5$, from bottom to top. The
upper and lower dashed lines are $2T/(\gamma \nu k^2)$
and $T/(\gamma \nu k^2)$, respectively. The crossover between these asymptotes
at $\Delta t \sim t_w$ ($k_n \sim k_w $) is clear. (b) Asymptotic time-delay limit,
$\Delta t \gg t_L$, for $t_w = 1,\,10,\,10^2,\,10^3,\,10^4$, and
$10^5$, from bottom to top. Dashed lines as in (a).}
\end{figure}

\begin{figure}[!tbp]
\centerline{
\includegraphics[angle=-0,width=6cm,clip=true]{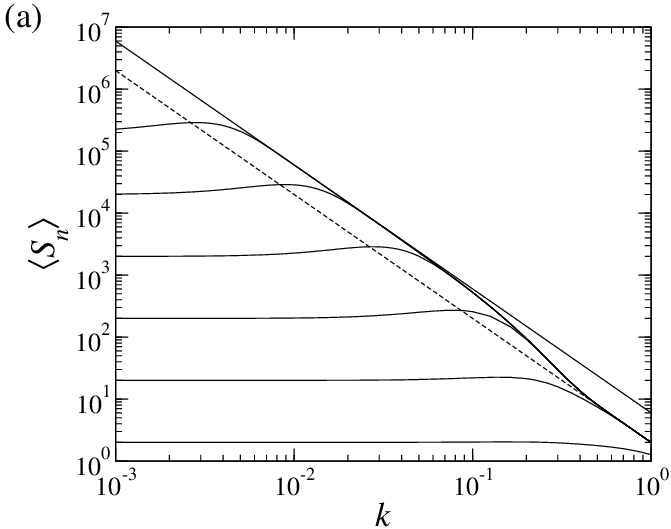}
\hspace{0.25in}
\includegraphics[angle=-0,width=6cm,clip=true]{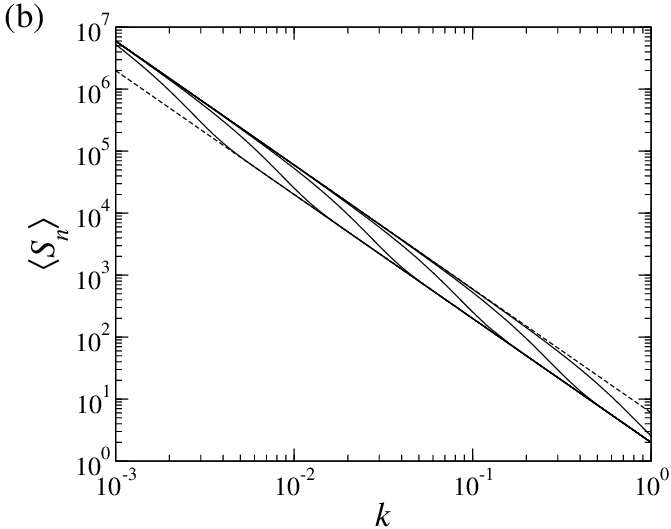}
}
\caption{\label{f:Sun_T0b} $T=1$ evolution of the structure factor
from an equilibrium initial condition at $T_0 =5>T$. (a) $t_w = 10$
and different time-delay values $\Delta t = 1,\,10,\,10^2,\,10^3,\,10^4$,
and $10^5$, from bottom to top. The dashed lines are $2T/(\gamma \nu k^2)$
and $(T_0+T)/(\gamma \nu k^2)$.  (b) Asymptotic time-delay limit,
$\Delta t \gg t_L$, for finite waiting-times
$t_w = 1,\,10,\,10^2,\,10^3,\,10^4$, and $10^5$,
from right to left. Dashed lines as in (a).}
\end{figure}

\begin{figure}[!tbp]
\centerline{
\includegraphics[angle=-0,width=6cm,clip=true]{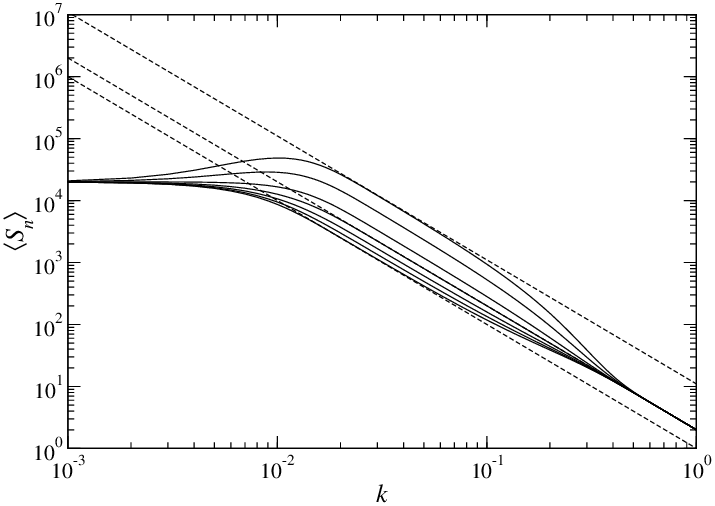}
}
\caption{\label{f:Sun_T0c} $T=1$ evolution of the structure factor for
different initial conditions, $T_0 = 0,\,0.2,\,0.5,\,1,\,2,\,5,\,10$,
from bottom to top. $t_w = 10$ and $\Delta t = 10^4$. The dashed lines
are $T/(\gamma\nu k^2)$ (middle) and
$(T+T_0)/(\gamma \nu k^2)$ with $T_0=0$ (bottom) and $T_0=10$ (top).}
\end{figure}

In Figs.~\ref{f:Sun_T00}-\ref{f:Sun_T0c} we display the analytic
results explained above. In Fig.~\ref{f:Sun_T00} we show the $T=1$
evolution of the structure factor $\langle S_n \rangle$ after heating
a perfectly flat initial condition, {\it i.e.} from $T_0=0$. The first panel illustrates the
time-delay saturation value and the power-law decay at large $k_n$,
where the change between the two asymptotes, which are proportional to
$T_0+T$ and $2T$, is clear. Figure~\ref{f:Sun_T00}~(b) shows the
crossover at $k_w\sim (\nu t_w)^{-1/2}$, between the two asymptotes
using different waiting-times and the large time-delay limit $\Delta t \gg t_L$. In this
case, {\it i.e.} $T>T_0$, the equilibrated asymptote, proportional to $2T$,
is the upper dashed line. In Fig.~\ref{f:Sun_T0b} (a) we show the
dynamics after a quench from $T_0>T$; the approach to the saturation
value at small $k_n$ is non-monotonic. In the asymptotic time-delay
limit shown in panel (b), $\langle S_n
\rangle$ crosses over from a higher value asymptote at small wave-vector to a
lower value at large wave-vector, proportional to $2T$ and corresponding to
equilibration, at the waiting-time dependent value $k_w$
in equation~(\ref{eq:scaling-S}).  In Fig.~\ref{f:Sun_T0c} we study
the dependence on the initial condition, $T_0$, finding that the
height and width of the bump increase with $T_0$.

\section{Two-times correlation length}
\label{sec:ell}

A practical definition of a growing two-times correlation length,
$l(\Delta t,t_w)$, is given by the fact that it marks the transition
between the saturation regime at small $k_n$ and the $1/k_n^2$
behavior at large $k_n$:
\begin{equation}
\label{eq:condcorr}
\langle S_{n=1} \rangle(\Delta t,t_w)=\lim_{\nu k_n^2 \Delta t\gg 1}
\langle S_{n=L/l} \rangle(\Delta t,t_w).
\label{eq:ell-def}
\end{equation}
This is similar to the numerical study in~\cite{Kolton} for {\it disordered}
elastic lines that, however, focuses only on a flat initial condition
and does not take into account the waiting-time dependence. In that
case, for $t_w=0$ and $T_0=0$, a crossover from a power law growth to a logarithmic growth was found, which is
related to the disordered energy landscape. In our case, without disorder,
we found a waiting time crossover between two regimes marking the separation between
\textit{non-equilibrated} and \textit{equilibrated} modes, as detailed in the following subsections.

\begin{figure}[!tbp]
\centerline{
\includegraphics[angle=-0,width=6cm,clip=true]{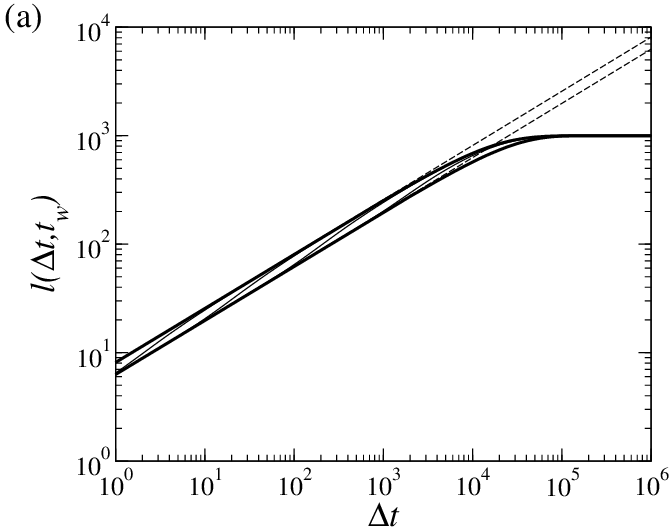}
\hspace{0.2in}
\includegraphics[angle=-0,width=6cm,clip=true]{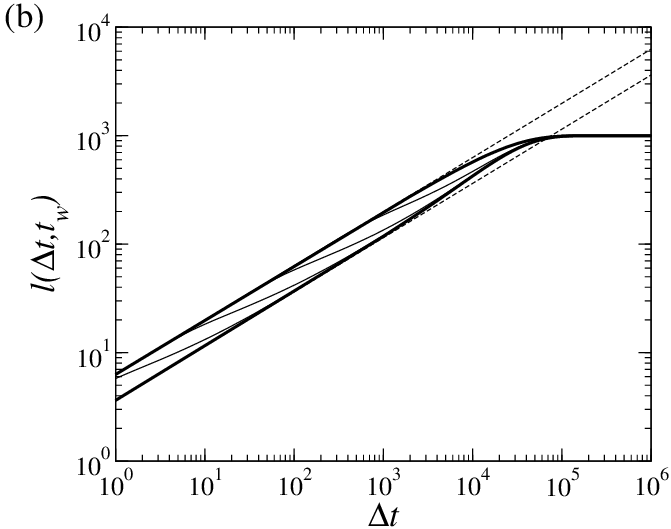}
}
\caption{\label{f:ell_EW_ttw_b} Aging of the two-times correlation
length $l(\Delta t,t_w)$ for $L=1000$. (a) $T_0=1 < T=5$ and (b)
$T_0=5 > T=1$. The continuous curves are the full solution with
$t_w=1,\,10,\,10^2,\,10^3$ from left to right in (a) and from bottom
to top in (b).  The thick curves correspond to $\lim_{t_w\gg t_L} l$
and $\lim_{t_w\ll t_L} l$; the former is located below for $T>T_0$ and
above for $T<T_0$. The dashed straight lines correspond to the
infinite size limit $L\rightarrow \infty$ taken at the outset and
describe the $\Delta t^{1/2}$ growth law.}
\end{figure}

\subsection{Stationary limit.}

In the stationary limit $t_w\gg t_L$ one finds
\begin{equation}
\lim_{t_w\gg t_L}
l(\Delta t,t_w)=L\sqrt{1-e^{-\Delta t/2 t_L}}.
\label{eq:ell-twinfty}
\end{equation}
This expression varies from $\lim_{\Delta t \ll t_L} \lim_{t_w\gg
t_L}l(\Delta t,t_w)=\sqrt{4 \pi^2 \nu \Delta t}$ in the growing regime
to $\lim_{\Delta t\gg t_L} \lim_{t_w\gg t_L} l(\Delta t,t_w)=L$ in the
saturation regime.  These limits correspond to the increases of the
saturation value of the structure factor and the limit in which all the
modes are equilibrated at the working temperature, respectively.  The
saturation time in the stationary regime corresponds to the time-delay
when these limits are equal, $\Delta t^*_{t_w\gg t_L}=2 t_L$, and it
is independent of $T$ and $T_0$.

\subsection{Decay of the initial condition.}

Setting $t_w=0$ (or, more generally, $t_w \ll t_L$) one finds
\begin{equation}
\lim_{t_w\ll t_L}
l(\Delta t,t_w)
=L
\sqrt{\frac{
T_0 \left(1- e^{-\Delta t/2 t_L}\right)^2 +
T \left( 1-e^{-\Delta t/t_L} \right)
}{
T_0+T
}
},
\label{eq:ell-tw0}
\end{equation}
which varies between $\lim_{\Delta t\ll t_L} \lim_{t_w\ll t_L} l(\Delta
t,t_w)=\sqrt{8 \pi^2 \nu \Delta t T/(T_0+T)}$ in the growing regime and $\lim_{\Delta
t\gg t_L} \lim_{t_w\ll t_L} l(\Delta t,t_w)=L$ in the saturation regime.  Equating these limits one finds the
saturation time associated to the growing correlation length $
\Delta t^*_{t_w=0}= \left(T_0/T+1\right)\,t_L$,
which depends on the initial condition.

\subsection{Aging scaling.}

In between the initial and the late stationary growths the two-times
correlation length ages; {\it i.e.} it depends also on $t_w$.
Equation~(\ref{eq:condcorr}) can be recast in a way that makes the scaling
solution apparent
\begin{equation}
\frac{l^2}{t_w} \left[ \left( \frac{T_0}{T} -1 \right) e^{-4 \pi^2 \nu \, t_w/l^2
}
+ 2 \right]
= g\left(\frac{T_0}{T}, \frac{\Delta t}{t_L}, \frac{t_w}{t_L}
\right).
\end{equation}
Indeed, this equation has a unique solution for $l^2/t_w$ for each set
of parameters in the right-hand-side.  In the growing regime, $\Delta t \ll t_L$, for $T_0>T$ the two-times
length $l$ moves from the upper asymptote, that corresponds to the
$t_w=0$ form in equation~(\ref{eq:ell-tw0}), to the lower one,
that corresponds to the $t_w\gg t_L$ form in equation~(\ref{eq:ell-twinfty}), at a
$\Delta t$ that increases with $t_w$. For $T_0<T$ the trend reverses:
$l$ moves from the lower asymptote to the upper one, representing the
$t_w\ll t_L$ and $t_w\gg t_L$ limits, respectively.  The crossover between
the two asymptotes occurs at a $t_w$-dependent $\Delta t$. More
precisely, in the growing regime one has
\begin{eqnarray}
l^2(\Delta t, t_w,T,T_0) = \left\{
\begin{array}{ll}
c_\infty \; \Delta t
\qquad & \Delta t \ll t_w
\\
2c_\infty T/(T_0+T) \; \Delta t
\qquad & \Delta t \gg t_w
\end{array}
\right.
\end{eqnarray}
with $c_\infty=4\pi^2 \nu$. It is clear that the relative value of the
prefactors depends on $T_0 > T$ or $T_0<T$.
The prefactor $c_0\equiv 2c_\infty/(T_0/T+1)$
vanishes at $T/T_0 \ll 1$, equals $c_\infty$ at $T_0=T$ and approaches
$2c_\infty$ at $T/T_0 \gg 1$.
The behaviour of the two-times dependent correlation length is summarized in
Fig.~\ref{f:ell_EW_ttw_b} which displays the waiting-time dependent
growth and subsequent saturation regime for two initial conditions
(a) $T_0<T$ and (b) $T_0>T$.

One should notice that $l$ increases monotonically with $\Delta t$ and
$t_w$ before reaching saturation for both $T_0>T$ and $T_0<T$. This
increase displays a multiplicative aging scaling behaviour between two
asymptotes with opposite trend depending on $T_0>T$ or $T_0<T$. These
kind of results where obtained, with additive aging scaling, in the
out-of-equilibrium relaxation of mixtures of soft spheres and
Lennard-Jones particles~\cite{Parisi} and the $3d$ EA
spin-glass~\cite{Chamon-etal,Ludovic} after a quench from a high
temperature (although the finite size saturation was not reached in
these numerical studies).  The heating case was not considered in
these models.

\section{Two-times roughness}
\label{sec:roughness}

The statics and equilibrium dynamics of elastic manifolds is usually
understood in terms of the time, temperature and system size dependence of their
averaged roughness or
width~\cite{Barabasi-Stanley}.  In most studies of interface dynamics
one compares the time-dependent configuration to the initial state,
typically taken to be perfectly flat (equilibrium at zero
temperature).  The thermal averaged (one-time dependent) roughness is then
\begin{equation}
\langle w^2 \rangle(t) =
L^{-d} \int d^d z \;
\langle [x(\vec z,t)-x(\vec z,0)]^2 \rangle
\; ,
\label{eq:roughness-def}
\end{equation}
where $x$ is the height of the surface and $\vec z$ is the position in
the $d$-dimensional substrate typically with cubic geometry and linear
size $L$.

The initial grow of the roughness is linear with time,
\begin{equation}
\lim_{\Delta t \to 0}  \langle w^2 \rangle(t) = 2Tt,
\end{equation}
which corresponds to a normal diffusion regime in which the beads on
the line are still uncorrelated. Note that there is no ballistic
regime since the EW equation represents overdamped Langevin
dynamics. However, these inertial effects could be relevant in polymer
-- molecular -- dynamics studies.  The normal diffusion regime can be
considered as a transient before the correlated dynamics of the
interface is reached. In the {\it single particle} regime the
roughness does not age. In the following we concentrate on the
correlated aging dynamics of the line.

The thermal (and disorder averaged if random interactions are present)
roughness follows the Family-Vicsek scaling \cite{Favi}, which means that it
crosses over from growth to saturation at $t_x\sim
L^{z}$~\cite{Barabasi-Stanley}:
\begin{equation}
\langle w^2 \rangle(t)
\sim L^{\zeta} f(t/t_x)
\; ,
\label{eq:tau}
\end{equation}
where the scaling function obeys $f(y) \sim y^\beta$ for $y \ll 1$
and $f(y) \sim 1$ for $y \gg 1$, with $\zeta$, $\beta$ and $z=\zeta
/ \beta$ the roughness, growth and dynamic exponents,
respectively. For the Edwards-Wilkinson (EW) line in $1+d$ dimensions
$\zeta=2-d$,
 $\beta=1-d/2$ and $z=2$.
In the presence
 of disorder, $\zeta$ is expected to take a
`thermal' value,
 $\zeta_{th}$ for $L<L_c(T)$ and a larger `disorder'
dominated value,
 $\zeta_{dis}$ for $L>L_c(T)$, both exponents being
$T$-independent~\cite{Barabasi-Stanley}. The other exponents, $\beta$
and $z$ may depend on $T$~\cite{schehr05} or even logarithmic
time-dependencies may exist~\cite{Kolton}.

We wish to take into account the waiting-time $t_w$ and consider also
more general initial conditions. To this end we generalize the
definition in equation~(\ref{eq:roughness-def}) to
\begin{equation}
\langle w^2 \rangle(t,t_w)
\equiv
\frac{1}{L} \int_0^L dz \;
\langle
\left[
\delta x(z,t) - \delta x(z,t_w)
\right]^2
\rangle
=\frac{2}{L} \sum_{n=1}^\infty \langle S_n \rangle(t,t_w)
\label{eq:roughness-def-gen}
\end{equation}
with $\delta x(z,t)$ defined in equation~(\ref{eq:defFouriermodes})
and especialized to $d=1$.
Note that the zero mode is not included in the sum.

In general,
one expects the dynamics to become stationary after an
equilibration time $t_L$; the generalized thermal averaged roughness
should then scale as in equation~(\ref{eq:tau}) with $t$ replaced by
$\Delta t$.  For not too short $L$, $t_L$ may
become very long and the dynamics might remain non-stationary with
$\langle w^2\rangle$
 depending on $t_w$ explicitly for $t_w<t_L$.
In~\cite{Bustingorryetal} we conjectured that the scaling of the
roughness in the {\it non-equilibrium relaxation} of {\it infinitely} long
elastic lines with or without quenched disordered potentials follows
the law
\begin{eqnarray}
&&
\langle w^2 \rangle(\Delta t,t_w)
\sim
\ell^\zeta(t_w) \; {\cal F}\left[\frac{\ell(t)}{\ell(t_w)}\right]
\label{eq:fitting0}
\end{eqnarray}
with $\ell(t)$ a growing length (dimensions are restored by prefactors
that we omit) and ${\cal F}$ a scaling function. For each waiting-time
this form approaches a stationary growth regime $\langle w^2 \rangle
\sim \ell^\zeta(\Delta t)$ when $t_w \ll \Delta t\ll t_L$ if ${\cal
F}(y) \sim y^{\zeta}$ for $y\gg 1$.  
It is also reasonable to assume ${\cal F}[\ell(t)/\ell(t_w)] \sim
\ell^\zeta(\Delta t)$ for $\Delta t\ll t_w$, that leads to a stationary
growth of the averaged roughness at very short time-delays.  This
result is explicitly realized in the power law case.  One may extend
this conjecture to apply to manifolds with internal dimension $D$ in a
transverse space with dimension $d$. The functional form of the
growing length, $\ell$, the scaling function, ${\cal F}$, and the
values of the exponent are expected to vary from case to case.

In a series of numerical studies one established that, in the
numerically accessible times, the quenched dynamics of a {\it
disordered} $1+1$ lattice model~\cite{Yosh98,Bustingorryetal}
satisfies the scaling in equation~(\ref{eq:fitting0}) with $\ell(t)\sim
t^{\alpha/\zeta}$ and $\alpha/\zeta$ a rather {\it small} exponent.  A
crossover to a logarithmic time-dependence~\cite{Kolton} is not
excluded although it was not seen in the out-of-equilibrium
relaxation.  It was shown that the roughness ages, by crossing over
between two asymptotes, for $\Delta t \gg t_w$ and $\Delta t \ll t_w$,
thus having
\begin{equation}
\langle w^2\rangle \sim  c_{1,2}(T) \, \Delta t^{\alpha(T)}
\; ,
\;\;\;\;
\mbox{with}
\;\;
\alpha(T) < \beta_{EW} = 0.5
\; ,
\label{eq:straight}
\end{equation}
and different proportionality constants. The waiting-time dependence
appears in the way these asymptotes are approached.  $\alpha(T)$ is a
generalization of the growth exponent, $\beta$, in surface growth
literature, and $\alpha(T)< \beta_{EW}$, with $\beta_{EW}$ the
Edwards-Wilkinson value, reflects that quenched
disorder slows down the dynamics.

In order to better understand this aging behaviour, trying to separate
the effects due to disorder from those related to the intrinsic
elastic character of the line, we present here results for the EW case
in $1+1$ dimensions {\it without disorder}. In this case the two-times
averaged roughness is given by
\begin{eqnarray}
\label{eq:w2Dttw}
\langle w^2 \rangle(\Delta t,t_w) = \frac{6w^2_0}{\pi^2} \;
\sum_{n=1}^{\infty}b_n(\Delta t,t_w)  + \frac{6w^2_\infty}{\pi^2} \;
\sum_{n=1}^{\infty}a_n(\Delta t,t_w),
\\
n^2 a_n(\Delta t,t_w) = 2 \left( 1-e^{-n^2 \frac{\Delta t}{2 t_L}} \right)
\left( 1-e^{-n^2 \frac{t_w}{t_L}} \right) + \left( 1-e^{-n^2 \frac{\Delta
t}{t_L}} \right) e^{-n^2 \frac{t_w}{t_L}},
\label{eq:an}\\
n^2 b_n(\Delta t,t_w) = \left( 1-e^{-n^2 \frac{\Delta t}{2 t_L}} \right)^2
e^{-n^2 \frac{t_w}{t_L}}, \label{eq:bn}
\end{eqnarray}
where we used the definitions
\begin{equation}
w^2_0\equiv T_0 L/(12 \gamma \nu), \qquad
w^2_{\infty}\equiv T L/(12 \gamma \nu).
\label{eq:def-winfty}
\end{equation}
The averaged two-times roughness can be written in terms of the scaled times
\begin{equation}
\langle w^2 \rangle(\Delta t,t_w)
=
\langle w^2 \rangle\left(\frac{\Delta
t}{t_L},\frac{t_w}{t_L}\right)
=
\langle w^2 \rangle\left(\frac{\Delta
t}{t_L},\frac{\Delta t}{t_w} \right)
\end{equation}
while temperatures appear separately, through $w_0^2$ and
$w_\infty^2$.  We first focus on the time-dependence of the roughness
of infinite lines, $L\to\infty$, before testing the dynamics at
different asymptotic limits, and we later reverse the order of limits
by considering finite lines, $L<\infty$.  We display the analytic
results in Figs.~\ref{f:w2_EW_ttw_a}-\ref{f:w2_EW_ttw_b1} in which we
approximate the series in the analytic expressions by using finite
sums, $\sum_{n=0}^\infty \rightarrow \sum_{n=0}^M$, with $M=1000$ in
all cases, except for lines with $L=3000$ for which we used $M=500$.

\subsection{Infinite system size.}

By taking $L\to\infty$ at the outset $t_L$ diverges and
\begin{equation}
\label{eq:w2LinfDttw0} \fl \lim_{L\to\infty} \langle w^2\rangle(\Delta t,t_w)
= \sqrt{\frac{2 t_w}{\pi \gamma^2\nu}} \left[ \left( T-T_0 \right) \left( 1 +
\sqrt{\frac{\Delta t}{t_w}+1} - \sqrt{\frac{2\Delta t}{t_w}+4} \right) +2T
\sqrt{\frac{\Delta t}{2 t_w}} \right],
\end{equation}
which can be written in the scaling form
\begin{equation}
\lim_{L\to\infty} \langle w^2 \rangle(\Delta t,t_w)
=
t_w^{1/2} \;
\widetilde{w}^2
\left(\frac{\Delta t}{t_w}\right),
\end{equation}
and admits the scaling form in equation~(\ref{eq:fitting0}) with $\ell(t) \sim t^{1/2}$.
Comparing now $\Delta t$ to a long waiting-time, that is to say taking $\Delta t\ll t_w$, one
finds that the waiting-time
dependence determines the crossover between two square-root dependencies
in  $\Delta t$ with different, temperature-dependent, prefactors:
\begin{eqnarray}
&&
\langle w^2 \rangle(\Delta t,t_w)
=
\left\{
\begin{array}{ll}
c^{w^2}_\infty(T) \; \Delta t^{1/2} \qquad & \Delta t \ll t_w
\\
c^{w^2}_0(T,T_0) \; \Delta t^{1/2} \qquad & \Delta t \gg t_w
\end{array}
\right.
\label{eq:w2Linfty}
\end{eqnarray}
with $c^{w^2}_0=[T+T_0(\sqrt{2}-1)]\sqrt{2/(\pi\gamma^2\nu)}$ and
$c^{w^2}_\infty=2T/\sqrt{\pi\gamma^2\nu}$. Note that
$c^{w^2}_0<c^{w^2}_\infty$ for $T>T_0$, the two constants are
identical at $T=T_0$, and $c^{w^2}_0>c^{w^2}_\infty$ for $T<T_0$.  The
two trends are shown in Fig.~\ref{f:w2_EW_ttw_b}~(b).  These results
are of the generic form proposed in~\cite{Bustingorryetal}, see
equation~(\ref{eq:fitting0}), with $\alpha=\beta_{EW}=1/2$ independently of
temperature, $\ell(t) \sim t^{1/2}$, and temperature dependencies of
the constants made explicit.

\subsection{Finite lines}

For finite $L$, by considering the long waiting-time limit $t_w\gg
t_L$ the stationary Family-Vicsek scaling form (\ref{eq:tau}) is
recovered.  Indeed, one has that
\begin{equation}
\lim_{t_w\gg t_L}
\langle w^2 \rangle(\Delta t,t_w)
=
w^2_{\infty} \;
\sum_{n=1}^{\infty} \frac{12}{\pi^2 n^2}
\left( 1-e^{-n^2 \Delta t/2 t_L} \right).
\end{equation}
This form for the stationary roughness contains the growing regime where
$
\lim_{\Delta t\ll t_L} \lim_{t_w\gg t_L} \langle w^2 \rangle
= 
c_\infty^{w^2}(T) \Delta t^{1/2}
$
as in eq.~(\ref{eq:w2Linfty}), and later crosses over to the saturation value
\begin{equation}
\lim_{\Delta t\gg t_L} \lim_{t_w\gg t_L}
\langle w^2\rangle(\Delta t,t_w)= 2 w^2_{\infty}=\frac{TL}{6 \gamma \nu}.
\label{eq:limit2}
\end{equation}

On the other hand, the averaged roughness at $t_w=0$ for finite $L$, or more precisely for $t_w \ll t_L$, also
follows the familiar Family-Vicsek scaling form (\ref{eq:tau}). In this case, the roughness crosses over from a growing regime
$
\lim_{\Delta t\ll t_L} \lim_{t_w\ll t_L} = c_0^{w^2}(T,T_0) \Delta t^{1/2}
$
to a saturation value which depends on the initial condition,
\begin{equation}
\lim_{\Delta t\gg t_L} \lim_{t_w \ll t_L}
\langle w^2 \rangle(\Delta t, t_w)
=w^2_0+ w^2_{\infty}=
\frac{(T_0+T)L}{12 \gamma \nu}.
\label{eq:limit1}
\end{equation}
Note that the saturation value reached in the stationary regime, after
having taken $t_w \gg t_L$, is {\it twice} the saturation
value obtained from the flat initial condition $T_0=0$. This result demonstrates how
important it is to be careful with the choice of $t_w$ to ensure that
one has reached the stationary regime when using numerical
simulations.

\begin{figure}[!tbp]
\centerline{
\includegraphics[angle=-0,width=6cm,clip=true]{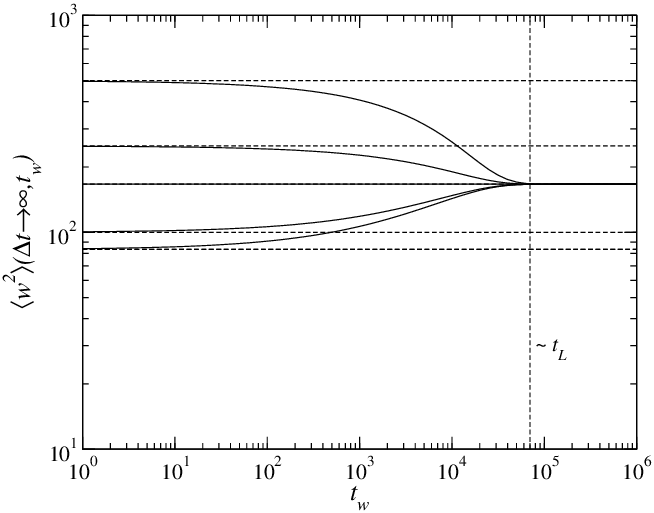}
}
\caption{\label{f:w2_EW_ttw_b1} Aging of the roughness at saturation,
$\lim_{\Delta t\gg t_L} w^2(\Delta t,t_w)$. $T=1$ and
$T_0=0,\,0.2,\,1,\,2,\,5$ from bottom to top. Dashed lines correspond
to $(T+T_0)L/(12 \gamma \nu)$. }
\end{figure}

When considering all the waiting-time dependence the roughness
interpolates between the different limiting values given above. For
sufficiently large but finite systems it is possible to find a well
defined aging of the growing regime when both $\Delta t$ and $t_w$ are
smaller than the saturation time. In this case one recovers the
scaling behaviour found in the infinite size system,
equations~(\ref{eq:w2LinfDttw0}) and (\ref{eq:w2Linfty}). For finite $L$
though this scaling form terminates at a characteristic time-delay,
$t_x(t_w)\propto t_L$ -- see below -- and $\langle w^2 \rangle$ later
saturates at a waiting-time-dependent value, which interpolates
between the two limiting saturation values, equations~(\ref{eq:limit2}) and
(\ref{eq:limit1}). These results are the same as the ones obtained
using the reverse order of time limits, see (\ref{eq:limit2}) and
(\ref{eq:limit1}). The $t_w$-dependence in the saturation regime
is shown in Fig.~\ref{f:w2_EW_ttw_b1}.

\begin{figure}[!tbp]
\centerline{
\includegraphics[angle=-0,width=6cm,clip=true]{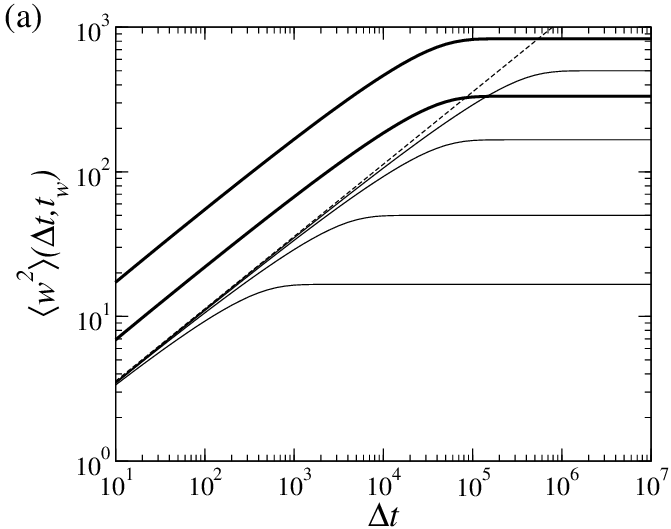}
\hspace{0.2in}
\includegraphics[angle=-0,width=6cm,clip=true]{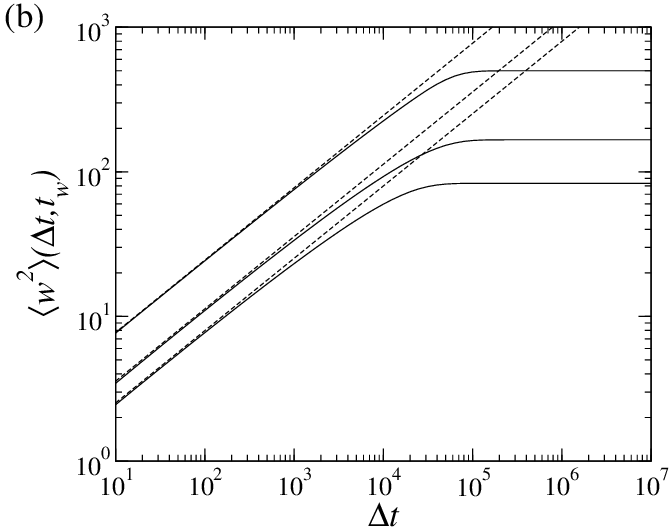}
}
\caption{\label{f:w2_EW_ttw_a} (a) Stationary roughness ($t_w
\gg t_L$) as a function of $\Delta t$. Thin lines
correspond to different system sizes $L=100,\,300,\,1000,\,3000$, from
bottom to top all at $T=1$. Thick lines
correspond to $L=1000$ and $T=2$ (bottom) and $T=5$ (top). Note that
the saturation time $t_x$ does not depend on $T$. The dashed line is
the limit $L \rightarrow \infty$ for $T=1$. (b) Roughness with $t_w=0$, $T=1$,
and different initial conditions corresponding, from bottom to top, to
$T_0=0,\,1$, and $5$. In the $t_w \gg t_L$ limit, the
roughness goes to its stationary solution, also given by the curve in
the middle.  Dashed lines correspond to $\lim_{L \rightarrow \infty}
\langle w^2 \rangle$
given by equation~(\ref{eq:w2Linfty}).}
\end{figure}

\begin{figure}[!tbp]
\centerline{
\includegraphics[angle=-0,width=6cm,clip=true]{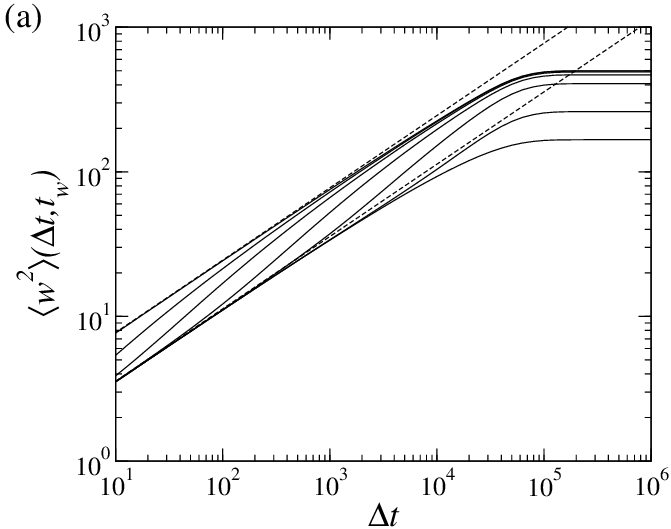}
\hspace{0.2in}
\includegraphics[angle=-0,width=6cm,clip=true]{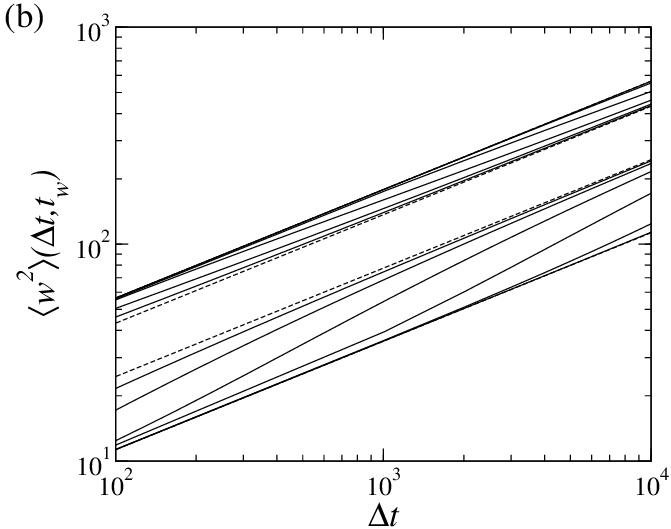}
}
\caption{\label{f:w2_EW_ttw_b} Aging of the two-times roughness,
$t_w=0,\,10^0,\,10^1,\,10^2,\,10^3,\,10^4,\,10^5$.
(a) $L=1000$, $T_0=5$ and $T=1$, $t_w$ increases from top to bottom.
The upper and lower dashed lines
correspond to $L \rightarrow \infty$ with $t_w=0$ and $t_w\gg t_L$,
respectively. (b) $L \rightarrow \infty$; $t_w$
increases from top to bottom in the lower set of curves ($T_0=5$ and
$T=1$) and from bottom to top in the upper set of curves ($T_0=1$ and
$T=5$).}
\end{figure}

Although in the previous analysis we used $t_L$ as a reference time to
determine the long time-delay limit. One can be more precise and
define a waiting-time dependent saturation time, $t_x(t_w)$. Equating the growing and saturation behaviour of the roughness, one can extract the saturation times
at two extreme values of $t_w$:
\begin{eqnarray}
t_x(t_w \ll t_L) &=&
\frac{\pi}{288 \nu} \left[ \frac{T+T_0}{T+T_0(\sqrt{2}-1)}\right]^2 \; L^2,
\label{eq:txtw0}
\\
t_x(t_w \gg t_L) &=& \frac{\pi}{144 \nu} \; L^2.
\label{eq:txtwinfty}
\end{eqnarray}
Both results are proportional to $t_L$, and $t_x(t_w)$ interpolates between these two values.
This justifies the use of
$t_L$ as the reference time-delay before saturation; the prefactors
are finite for all $T$ and $T_0$, the former is larger (smaller) for
$T<T_0$ ($T>T_0$) and become identical at $T=T_0$.  The saturation
time and its dependence on $L$ are visible in
Figs.~\ref{f:w2_EW_ttw_a} and \ref{f:w2_EW_ttw_b} (a).

\subsection{Scaling relations}
\label{sec:scalrel}

Here we present the main results concerning the aging properties of the roughness.
We have shown for the aging regime that
\begin{eqnarray}
\begin{array}{lcll}
\;\;t_x &\sim& a(T) \; L^z, \qquad & z=2,
\nonumber\\
\langle w^2 \rangle_{\Delta t \gg t_x}
&\sim& L^\zeta, \qquad & \zeta=1,
\nonumber\\
\langle w^2 \rangle_{\Delta t \ll t_x} &\sim& c(T,T_0,\Delta t/t_w) \; t_w^\beta,
\qquad & \beta=1/2,
\end{array}
\end{eqnarray}
see equations~(\ref{eq:txtw0}) and (\ref{eq:txtwinfty}), equation~(\ref{eq:def-winfty})
and  equation~(\ref{eq:w2Linfty}), respectively. The prefactor
$c(T,T_0,\Delta t/t_w)$ approaches
$c_\infty^{w^2}(T) (\Delta t/t_w)^{1/2}$ and
$c_0^{w^2}(T,T_0) (\Delta t/t_w)^{1/2}$,
for $\Delta t \ll t_w$ and $\Delta t \gg t_w$, respectively.
These relations imply
\begin{equation}
t_x \sim a(T) (w_\infty^2)^{z/\zeta}
\sim \left[ a(T)^{\zeta/z} w_\infty^2 \right]^{z/\zeta}
\; ,
\end{equation}
where we have used that $\langle w^2 \rangle_{\Delta t \gg t_x} \sim w^2_\infty$ for simplicity.
From matching the end of the growth regime with saturation
at $t_w\ll \Delta t=t_x$
one has
\begin{equation}
c(T,T_0,\Delta t/t_w) \, t_x^\beta \sim c_0^{w^2}(T,T_0) \, t_x^\beta \sim w_\infty^2
\qquad \mbox{and} \qquad
\beta=\zeta/z,
\end{equation}
(the latter condition is satisfied by the values of the exponents found).
Thus one finally has that
\begin{equation}
a(T)^{-\zeta/z}=c_0^{w^2}(T,T_0).
\end{equation}
Note that the exponents take simple values in the EW case but, in general,
they can be $T$-dependent (not $\zeta$)~\cite{schehr05}.

\section{Mean-squared and center-of-mass displacements}
\label{sec:MSD}

The two-times roughness, equation~(\ref{eq:roughness-def-gen}),  may also be written as
\begin{eqnarray}
\langle w^2\rangle(t,t_w) &=
 \langle B\rangle(t,t_w) - \langle D\rangle(t,t_w)
\nonumber\\
&=
\left\langle
\overline{\left[ x(z,t) - x(z,t_w) \right]^2} \right\rangle -
\left\langle \left[ \overline{x(t)} - \overline{x(t_w)} \right]^2
\right\rangle,
\end{eqnarray}
where $\langle B\rangle$ and $\langle D\rangle$ represent
the averaged mean-squared-displacement of the differential line segments
and the center of mass of the line, respectively, see
Fig.~\ref{fig:sketch-lines}.
This relation simply states that the roughness is a measure of the fluctuations around the center of
mass of the line.

It is simple to show that the center of mass diffuses normally.
Indeed, integrating equation~(\ref{eq:EW}) over the line length one has
$
\partial_t \overline{x}(t)=\xi'(t),
$
with
$\xi'(t)=\frac{1}{L}\int_0^L dz\;\xi(z,t)$,
$\langle \xi'(t) \rangle =0$,
$\langle \xi'(t) \xi'(t') \rangle = \frac{2T}{\gamma L} \delta(t-t')$,
and
\begin{equation}
\label{eq:Dcenmass}
\langle D\rangle(t,t_w)=
\langle D\rangle(\Delta t) =
\frac{2T}{\gamma L} \Delta t.
\end{equation}
The diffusion constant is an inverse function of the line length.

In previous studies of the elastic line out-of-equilibrium
dynamics~\cite{Yosh98,Cule} one focused on the
mean-squared-displacement that, in the case of the EW line, is just
given by $\langle B\rangle(t,t_w)=\langle w^2\rangle(t,t_w)+2T/(\gamma
L) \; \Delta t$.  At short time-delay $w^2$ and $B$ are
practically identical while in the saturation regime the displacement
is just given by the normal diffusion law.

\section{The incoherent scattering function}
\label{sec:incoherent}

\begin{figure}[!tbp]
\centerline{
\includegraphics[angle=-0,width=6cm,clip=true]{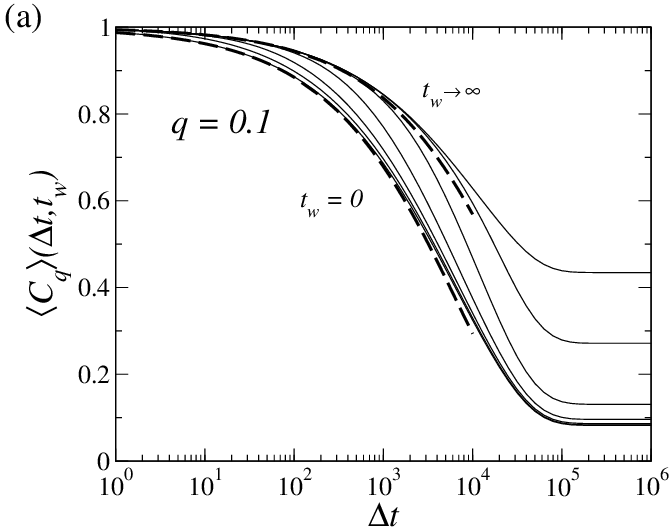}
\includegraphics[angle=-0,width=6cm,clip=true]{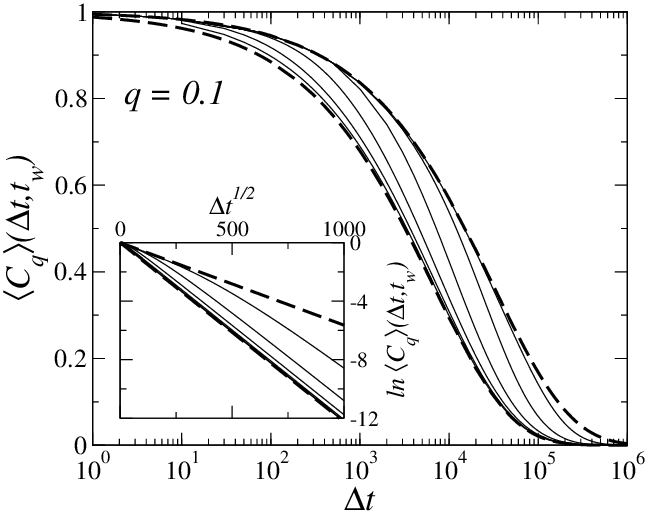}
}
\caption{The wave-vector dependent correlation function defined in
equation~(\ref{eq:Cqdef}) and given in terms of $\langle \, w^2\, \rangle$
by equation~(\ref{eq:CqGauss}) for a Gaussian process.  $\langle C_q
\rangle$ at $T_0=5$, $T=1$, $q=0.1$ and $t_w =0, \, 1, \, 10, \, 10^2,
\, 10^3, \, 10^4$, and $10^5$.  (a) $L=1000$.  The thick dashed lines
correspond to the limits $L \rightarrow \infty$ with $t_w=0$ (lower
curve) and $t_w \gg t_L$ (upper curve).  (b) $L \rightarrow \infty$
limit; the rest of the parameters are as in panel (a).  The inset
shows the same curves in a different scale, {\it i.e.} $\ln \langle
C_q\rangle$ vs.  $\Delta t^{1/2}$.  }
\label{f:corrq}
\end{figure}

\begin{figure}[!tbp]
\centerline{
\includegraphics[angle=-0,width=6cm,clip=true]{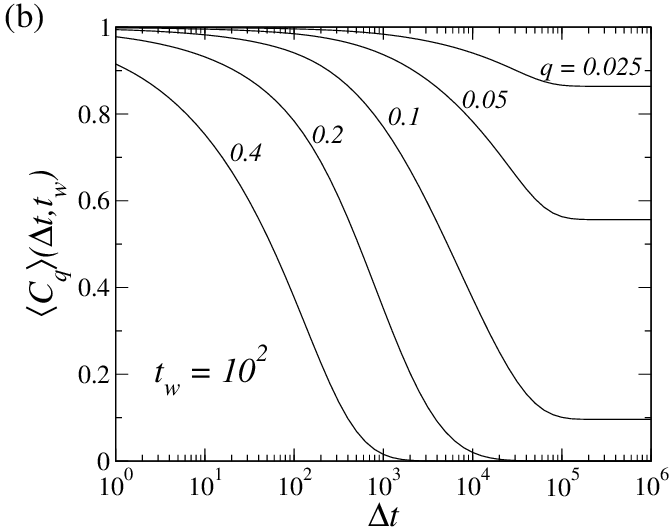}
}
\caption{$\langle C_q\rangle$ for $L=1000$, $t_w=10^2$ and
different wave-vectors $q$ as indicated. $T_0=5$ and $T=1$.}
\label{fig:Cq-q}
\end{figure}

The dynamics of glassy systems is usually analyzed
in terms of the wave-vector dependent
incoherent scattering function:
\begin{equation}
\langle C_q \rangle(t,t_w) = N^{-1} \sum_{i=1}^N \langle \,
e^{i\vec q[\vec r_i(t)-\vec r_i(t_w)]} \, \rangle
\end{equation}
with $N$ the total number of particles, $\vec r_i(t)$ the 
time-dependent position of particle $i$ and $\vec q$ a wave-vector. 
$C_q$ is measured numerically and experimentally. In the context of
elastic lines, one defines
\begin{equation}
\langle C_q \rangle(\Delta t,t_w) = L^{-1} \int dz \; \langle \,
e^{i q [\delta x(z,t)-\delta x(z,t_w)]} \, \rangle.
\label{eq:Cqdef}
\end{equation}
In the EW case the displacement in the exponential is a Gaussian
random variable and
\begin{equation}
\langle C_q \rangle(\Delta t,t_w) = e^{-\frac{q^2}{2} \langle \, w^2 \, \rangle(\Delta t,t_w)}.
\label{eq:CqGauss}
\end{equation}
The incoherent scattering function $\langle C_q \rangle$ is simply related to the roughness, $\langle w^2
\rangle$, analyzed in Sect.~\ref{sec:roughness}.  Figure~\ref{f:corrq}
displays the time-delay decay of $ \langle C_q\rangle$ at $q=0.1$
using several waiting-times. For finite lines the saturation in
$\langle w^2\rangle$ is attained at sufficiently long $\Delta t$ and
the correlation reaches a waiting-time dependent plateau with its
height increasing with $t_w$, Fig.\ref{f:corrq}~(a). For sufficiently long lines the
saturation regime can be pushed beyond the observed time-delay window
as exemplified in the limit $L\to\infty$, Fig.~\ref{f:corrq}~(b). In Fig.~\ref{fig:Cq-q}
we display the dependence of $\langle C_q\rangle$ on $\Delta t$ for
fixed $t_w$ and different values of $q$. One notices that the small
$q$ correlations saturate in the $\Delta t$ window while large $q$
correlations relax to zero. This is to be expected since $q^2 w^2_\infty$
scales as $q^2L$ and the effect of decreasing
$q$ is like decreasing $L$.  Note the similarity between these plots
and light-scattering measurements in clay suspensions
(laponite)~\cite{laponite}.

\section{Response functions and FDT}
\label{sec:response}

We now compute the linear response function of several
observables related to the two-times correlations studied above.

\subsection{Center of mass response}

In order to evaluate the linear response we have to switch on a
perturbing field coupled to the observable of interest.
Let us start with the linear response function
associated with the center of mass diffusion, $\langle\chi^D\rangle$.
The effect of
a perturbing field, $h$, applied on the center of mass
after time $t_w$ is described by the  term
\begin{equation}
{\cal H}'^D=
-h \, L \overline{x}(t) \, \theta(t-t_w)=
-h \; \int_0^L dz\; x(z,t) \; \theta(\Delta t)
\end{equation}
that is added to the energy.
Calling $\overline{x}^h(t)$ and $\overline{x}(t)$
the center of mass position with and without field, respectively,
the linear response function is
\begin{equation}
\langle \chi^D \rangle(t,t_w)=\frac{1}{hL}
\left\langle \overline{x}^h(t)-\overline{x}(t) \right\rangle
\end{equation}
which depends only on the time difference and satisfies the
FDT for any $t$ and $t_w$,
\begin{equation}
\langle D \rangle(\Delta t)=2 T \langle \chi^D \rangle(\Delta t).
\end{equation}

\subsection{Roughness response}

\begin{figure}[!tbp]
\centerline{
\includegraphics[angle=-0,width=6cm,clip=true]{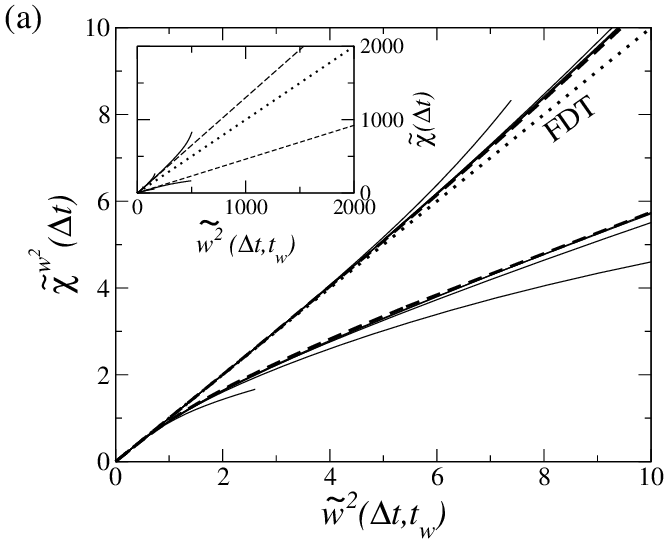}
\hspace{0.2in}
\includegraphics[angle=-0,width=6cm,clip=true]{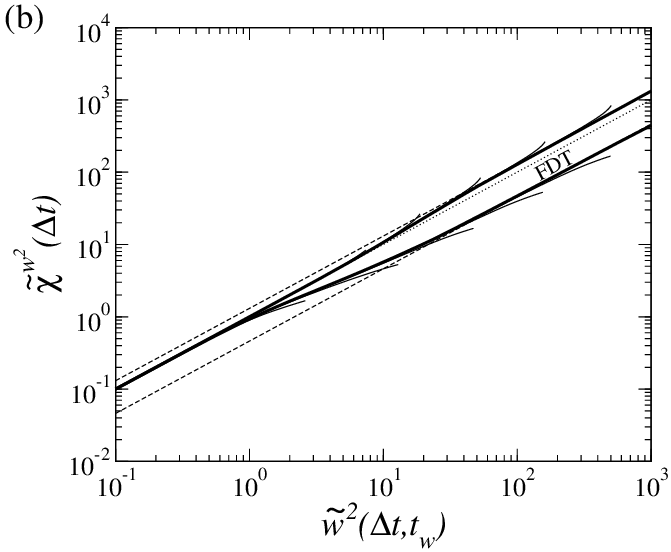}
}
\caption{\label{f:FDT_EW_ttw_a} Violation of the FDT in the aging
regime of the EW equation. The parametric plots
$\widetilde{\chi}^{w^2}= 1/(2T_{eff})\, \widetilde{w}^2$ are
constructed from the scaled variables defined by
$\widetilde{w}^2=t_w^{-1/2} \langle w^2 \rangle$ and
$\widetilde{\chi}^{w^2}=t_w^{-1/2} \langle \chi^{w^2} \rangle$. (a)
The parametric plot in linear scale showing the departure from the FDT
(dotted line). Upper and lower sets of curves correspond to
$T_0=1,\,T=5$ and $T_0=5,\,T=1$, respectively. Thick dashed lines
represent the $L \rightarrow \infty$ limit. Thin lines are for
$L=1000$ and different waiting-times, and show the finite size
signature in the FDT parametric plot. The inset shows the large scale
violation of FDT with two straight lines, which corresponds to
effective temperatures larger (for $T_0>T$) and smaller (for $T_0<T$)
than the working temperature $T$. The initial FDT regime is not
clearly observed in this scale. (b) The parametric plot in log-log
scale showing that the violation of the FDT is given by
equation~(\ref{eq:yTeffDtinf}) at $\Delta t \gg t_w$.}
\end{figure}

The energy contribution of a field  conjugated to the roughness is
\begin{equation}
{\cal H}'^{w^2}=-h \;
\int_0^L dz\; \left[ x(z,t)- \overline{x}(t) \right]
s(z)
\theta(\Delta t).
\end{equation}
As usual, $s(z)$ are {\it i.i.d.} quenched random variables
taking  values $s(z)=\pm 1$ with equal probability:
$\langle s(z) \rangle=0$ and
$\langle s(z) s(z')\rangle=\delta(z-z')$.
The associated response function is
\begin{eqnarray}
\langle \chi^{w^2}\rangle(t,t_w) &=
 \frac{1}{hL} \left\langle \int_0^L dz\;
\left[ \delta x^h(z,t) - \delta x(z,t) \right] s(z) \right\rangle
\nonumber \\
&=
 \frac{2}{h} \sum_{n=1}^\infty \left\langle \left[ c_n^h(t) - c_n(t)
\right] s_n \right\rangle, \label{eq:resp-w2}
\end{eqnarray}
with $s_n$ defined through
$\delta s(z) = s(z)-\overline{s}=\sum_{n=-\infty}^{\infty} s_n e^{i
k_n z}$.
Here $\langle \ldots \rangle$ also indicates the average over the $s_n$
distribution.  One finds
\begin{equation}
\langle \chi^{w^2} \rangle(\Delta t)
= \left( 1-e^{- \nu k_n^2 \Delta t} \right)
\end{equation}
where we used $\langle s_n^2 \rangle = 1/L$.
Interestingly enough, the linear response is stationary for all
$t_w$ while the roughness is not. Therefore, the FDT is not respected
for waiting-times $t_w \ll t_L$,
and its modification is discussed in Sect.~\ref{subsec:FDT}
and Fig.~\ref{f:FDT_EW_ttw_a}.
In the stationary limit
$t_w \gg t_L$ the roughness
becomes stationary and the FDT holds, {\it i.e.}
\begin{equation}
\label{eq:w2FDT}
\lim_{t_w\gg t_L} \langle w^2\rangle(\Delta t,t_w)= 2 T \langle
\chi^{w^2}\rangle(\Delta t).
\end{equation}
This statement also implies that the FDT does not hold for $t_w=0$ and $T=0$, pointing again that one should be careful with the choice of $t_w$ and the stationary limit.

\subsection{Mean-squared-displacement response}

The effect of a
perturbing field conjugated to the mean-squared-displacement
is represented by
\begin{equation}
{\cal H}'^B=-h \;\int_0^L dz\; x(z,t) s(z) \; \theta(\Delta t),
\end{equation}
with the random $s(z)$ distributed as above.
The linear response function is defined as
\begin{eqnarray}
\langle \chi^B \rangle(t,t_w) &=\frac{1}{hL}\left\langle \int_0^L dz\; \left[
x^h(z,t) - x(z,t) \right] s(z) \right\rangle
\nonumber \\
&=
 \langle \chi^{w^2} \rangle(t,t_w) + \frac{\langle \overline{s} \rangle}{h}
 \left[ \overline{x}^h(t) - \overline{x}(t) \right].
\end{eqnarray}
The last term of this expression represents the center-of-mass
response to a field of intensity $h'=h/\langle \overline{s} \rangle$.
Thus, in the long waiting-time limit, $\langle \chi^B\rangle$ is also
stationary and simply related to the roughness and center-of-mass
responses,
\begin{equation}
\langle \chi^B\rangle(\Delta t)=\langle \chi^{w^2}\rangle(\Delta t)+
\langle \chi^D\rangle(\Delta t).
\end{equation}
In this case, the FDT is not satisfied in general but in the stationary regime
it is:
\begin{equation}
\lim_{t_w\gg t_L}
\langle B\rangle(t,t_w)= \langle B\rangle(\Delta t) =
2 T \langle \chi^B \rangle(\Delta t).
\end{equation}
\begin{figure}[!tbp]
\centerline{
\includegraphics[angle=-0,width=6cm,clip=true]{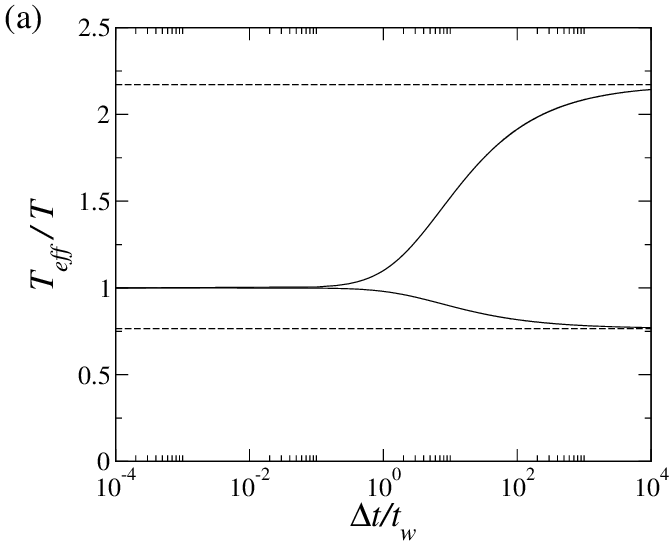}
\hspace{0.2in}
\includegraphics[angle=-0,width=6cm,clip=true]{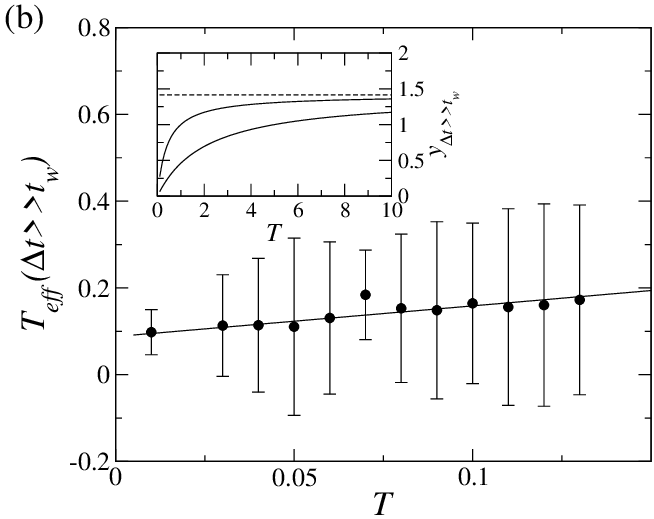}
}
\caption{\label{f:Teff_EW_ttw_a} The effective temperature $T_{eff}$
characterizing the violation of the FDT in the EW equation. (a)
Evolution with the rescaled time $\Delta t/t_w$ showing that FDT holds
when $\Delta t \ll t_w$ and it does not hold when $\Delta t \gg
t_w$. Upper and lower curves correspond to $T_0=5,\,T=1$ and
$T_0=1,\,T=5$, respectively. The dashed lines represent the limit
$\Delta t \gg t_w$ in equation~(\ref{eq:yTeffDtinf}). (b) $T_{eff}(\Delta t
\gg t_w)$ vs. $T$ for $T_0=0.3$ (continuous line) compared to the data
in Fig.~25 in~\cite{us} describing the effective temperature of
independent elastic lines (with the Josephson correction to the
elasticity) moving in a {\it quenched random environment}.  The inset
shows $y_{\Delta t \gg t_w}=\lim_{\Delta t \gg t_w} T/T_{eff}(T)$
as a function of temperature for $T_0=1,\,T=5$ (lower
curve) and $T_0=5,\,T=1$ (upper curve).  The dashed line is the limit
$T \gg T_0$.}
\end{figure}

\subsection{FDT and effective temperature}
\label{subsec:FDT}

\begin{figure}[!tbp]
\centerline{
\includegraphics[angle=-0,width=6cm,clip=true]{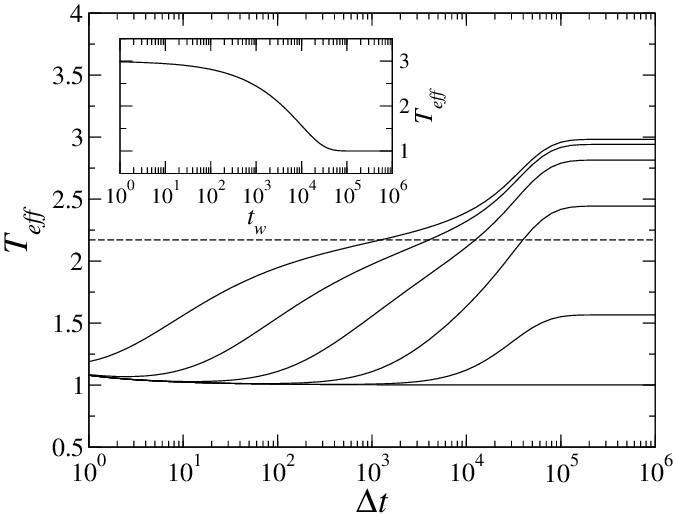}
}
\caption{\label{f:Teff_EW_ttw_last} The effective temperature $T_{eff}$
as a function of time-delay for $t_w=1,\,10,\, 10^2,\, 10^3,\, 10^4$
and $10^5$ from top to bottom; $L=1000$, $T_0=5$ and $T=1$.  The dashed line is
$\lim_{t_w\ll \Delta t} \lim_{L\to\infty} T_{eff}$ [the same dashed curve as
in Fig.~\ref{f:Teff_EW_ttw_a} (a)] and justifies the shoulder for
small $t_w$. The inset shows $T_{eff}(\Delta t=10^6 \gg t_L, t_w)$ as
a function of $t_w$; it is clear that at sufficiently long $t_w$ one
recovers $T_{eff}=T$.}
\end{figure}

Since FDT holds for $t_w \gg t_L$, it is interesting to study the
violation of the FDT at finite $t_w$. To this end we use the $L
\rightarrow \infty$ limit, where the roughness takes the scaling form
in equation~(\ref{eq:w2LinfDttw0}), and
\begin{equation}
\lim_{L\to\infty}
\langle \chi^{w^2}\rangle(\Delta t)
=
t_w^{1/2}
\widetilde{\chi}^{w^2} \left( \frac{\Delta t}{t_w} \right)
\;\;\;
\mbox{with}
\;\;\;
\widetilde{\chi}^{w^2} \left( \frac{\Delta t}{t_w} \right) =
\sqrt{\frac{1}{\pi \gamma^2 \nu} \; \frac{\Delta t}{t_w} }.
\label{eq:chiscaling}
\end{equation}
Once we eliminated the $t_w^{1/2}$ diffusive factor one can associate the
ratio between $\widetilde{w}^2$ and $\widetilde{\chi}^{w^2}$ with an
effective temperature~\cite{Yosh98,us},
\begin{equation}
\widetilde{\chi}^{w^2}=\frac{1}{2T_{eff}} \widetilde{w}^2.
\end{equation}
which, in this case, depends on $T$, $T_0$, $\Delta t$ and $t_w$.
From equation~(\ref{eq:w2LinfDttw0}) we obtain
\begin{equation}
\label{eq:yTeff}
T_{eff}= T \left[ 1+\frac{T-T_0}{T} \left(
\sqrt{\frac{t_w}{2 \Delta t}}+\frac{1}{\sqrt{2}}
\sqrt{1+\frac{t_w}{\Delta t}}-\sqrt{1+\frac{2 t_w}{\Delta t}} \right)
\right].
\end{equation}
In the case $T=T_0$ one recovers $T_{eff}=T$ as expected.
One can check that $T_{eff}>T$  or $T_{eff}<T$
whenever $T_0>T$ or $T_0<T$. In the two extreme cases $\Delta t\ll
t_w$ and $\Delta t\gg t_w$ one finds two waiting-time {\it independent}
values of $T_{eff}$:
\begin{eqnarray}
\label{eq:yTeffDtinf}
T_{eff}=
\left\{
\begin{array}{ll}
T \; \qquad &\Delta t\ll t_w
,\\
T+ (T_0-T) \left( 1-\frac{1}{\sqrt{2}}\right)
\; \qquad &\Delta t\gg t_w,
\end{array}
\right.
\end{eqnarray}
see Figs.~\ref{f:FDT_EW_ttw_a} and \ref{f:Teff_EW_ttw_a},
indicating that fast modes are equilibrated while the slow ones are not.
These results are very similar to what has been found numerically for the
dynamics of elastic lines in a quenched random potential~\cite{Yosh98}
and in models of {\it interacting}
elastic lines in quenched random environments that describe the vortex
glass in high $T_c$ superconductors~\cite{us}, see
Fig.~\ref{f:Teff_EW_ttw_a}. Note that the waiting-time dependence in
$T_{eff}$ only marks the crossover between the two asymptotic regimes.
At still longer $\Delta t$ such that $\Delta t\gg t_x\sim t_L$ the FDT
result, $T_{eff}=T$, is recovered. This is shown in
Fig.~\ref{f:Teff_EW_ttw_last}, a result that, once again, resembles
what was found in laponite~\cite{Abou}.

\section{Fluctuations}
\label{sec:fluctuations}

Upto this point we studied the scaling properties of several two-times
quantities averaged over the length of the line {\it and} thermal
noise. A more refined investigation of interface dynamics, currently
done theoretically and experimentally, deals with the fluctuating
quantities such as the distribution functions of the width of
heights~\cite{Racz}, the density of local maxima or minima of
heights~\cite{Satya,Greg}, the statistics of first passage times, {\it
etc.}  It is also clear by now that to gain a complete understanding
of glassiness one should also understand the dynamic
fluctuations~\cite{Chamon-etal,Chamon-Cugliandolo}. In the glassy
context one expects that the competition between different time or
length scales in the system reflects in the way the probability
distribution functions (pdfs) behave.

R\'acz proposed that, for elastic systems, the scaling form of the
distribution function characterizing the roughness fluctuations may
serve to classify the systems into different universality
classes~\cite{Racz}. For a given system size the pdf of the roughness, 
$P_L(w^2)$, scales as~\cite{static-inter}
\begin{equation}
w^2_\infty P_L(w^2) = \Phi\left( \frac{w^2}{w^2_\infty} \right),
\label{eq:PhiRacz}
\end{equation}
in the saturation regime, and the form of the scaling function was
shown to be useful to differentiate between the EW and KPZ
universality classes~\cite{Racz}. The full stationary pdf of the EW
roughness was also computed, showing the same scaling behavior as in
equation~(\ref{eq:PhiRacz}). These works considered only the $t_w=0$ case
with a flat initial condition ($T=0$).  Recently, a simulation study
of a disordered elastic-line model defined on the lattice demonstrated
that the scaling form of the distribution function must be modified to
include the aging effects~\cite{Bustingorryetal}.

In the body of this Section we analyze the thermal noise-induced dynamic
fluctuations of the two-times quantities defined previously during the
aging relaxation.

\subsection{Roughness distribution}

Rewriting the roughness $w^2$ in terms of  $u(z,t,t_w)=x(z,t)-x(z,t_w)$,
the pdf of the two-times roughness, $P_L(w^2)$, is given by
\begin{displaymath}
P_L(w^2)=\int {\cal D}[x]{\cal D}[x'] \; \delta \left[ w^2-
\left( \overline{u^2(t,t_w)}-\overline{u(t,t_w)}^2 \right) \right]
p\left( {x},t;{x'},t_w \right),
\end{displaymath}
where ${\cal D}[x]$ is the measure over
$x$ configurations; $x$ and $x'$ represent the configurations at time
$t$ and $t_w$, respectively; and $p\left( {x},t;{x'},t_w \right)$ is
their joint probability density. The Laplace
transform
$G_L(\lambda)=\int_0^\infty d \alpha \; P_L(\alpha) \;
e^{-\lambda \alpha}$,
can be written as the path integral
\begin{equation}
G_L(\lambda)=\int {\cal D}[x]{\cal D}[x'] \;
p\left( {x},t;{x'},t_w \right) e^{-\lambda \left( \overline{u^2(t,t_w)}-\overline{u(t,t_w)}^2 \right)}.
\end{equation}
Using the independent
Fourier modes defined in equation~(\ref{eq:defFouriermodes})
\begin{equation}
\label{eq:GFmodes}
\fl
G_L(\lambda,t,t_w)= {\cal N} \int \prod_{n=1}^\infty
dc_n\,dc_n^*\,dc'_n\,{dc'}_n^* \; p^2\left[ c_n(t),c_n(t_w)|c_n(0) \right] \;
e^{-2\lambda |c_n(t)-c_n(t_w)|^2},
\end{equation}
where ${\cal N}$ is a normalization factor ensuring
$G_L(0)= 1$ at all times.
The quantity $p\left[ c_n(t),c_n(t_w)|c_n(0) \right]$ is the joint
probability density of having $c_n$ at time $t$ and $c'_n=c_n(t_w)$
at time $t_w$, given the initial condition $c_n(0)$. This quantity can
be expressed as
$p\left[ c_n(t),c_n(t_w)|c_n(0) \right]=p\left[ c_n(t)|c_n(t_w) \right] p\left[ c_n(t_w)|c_n(0) \right]$,
where $p\left[ c_n(t)|c_n(t') \right]$ is the conditional probability
of evolving from $c_n(t')$ to $c_n(t)$ in the
time interval $t-t'$. This states that the joint
probability is simply the product of the probabilities of evolving
from the initial condition to the intermediate configuration
$c_n(t_w)$, and from there to the configuration $c_n(t)$, and
satisfies
\begin{equation}
p\left[ c_n(t)|c_n(0) \right]=\int dc'_n\,{dc'}_n^*\;
p\left[c_n(t),c_n(t_w)|c_n(0) \right].
\end{equation}
The conditional probability $p\left[ c_n(t)|c_n(t') \right]$ is
given by the complex Gaussian
\begin{equation}
p\left[ c_n(t)|c_n(t') \right] = \frac{1}{2 \pi \sigma_n^2(t-t')} \;
e^{- \frac{\left| c_n(t) - c_n(t') e^{-\nu k_n^2 (t-t')} \right|^2}{2
\sigma_n^2(t-t')}},
\end{equation}
where
\begin{equation}
\sigma_n^2(t-t')=\frac{T}{L \gamma \nu k_n^2} \left[ 1-e^{-2 \nu k_n^2
(t-t')} \right].
\end{equation}
Therefore, the joint probability function for the Fourier
modes at times $t$ and $t_w$ becomes
\begin{equation}
p\left[ c_n(t),c_n(t_w)|c_n(0) \right] =
\frac{e^{- \frac{\left| c_n(t)
- c_n(t_w) e^{-\nu k_n^2 \Delta t} \right|^2}{2 \sigma_n^2(\Delta t)}}
e^{ - \frac{\left| c_n(t_w) - c_n(0) e^{-\nu k_n^2 t_w}
\right|^2}{2 \sigma_n^2(t_w)}}}{(2 \pi)^2
\sigma_n^2(\Delta t) \sigma_n^2(t_w)}.
\end{equation}
After some algebra one finds the normalization factor
${\cal N}=\prod_{n=1}^\infty 16 \pi^2 \sigma_n^2(\Delta t)
\sigma_n^2(t_w)$
and
\begin{equation}
\label{eq:GDttw}
G_L(\lambda)
=
\prod_{n=1}^\infty
\frac{e^{-\frac{2 \lambda |c_n(0)|^2 \left( 1-
e^{-\nu k_n^2 \Delta t} \right)^2 e^{-2 \nu k_n^2 t_w}}{1+\lambda
w_\infty^2 a_n(\Delta t,t_w)}}}{1+\lambda w_\infty^2 a_n(\Delta t,t_w)}
,
\end{equation}
with the coefficients $a_n$ defined  in equation~(\ref{eq:an}).
This is the two-times generalization of the result in~\cite{Zoltan}, including arbitrary initial conditions.
The averaged two-times roughness follows from
$G_L$ as
$\langle w^2\rangle (\Delta t,t_w)=\left. -\partial_\lambda G_L(\lambda,t,t_w)
\right|_{\lambda=0}$,
which allows to recover the result in equation~(\ref{eq:w2Dttw}) for
the initial condition $|c_n(0)|^2=T_0/(L \gamma \nu k_n^2)$.

Since the two-times roughness pdf is given by
\begin{equation}
P_L(w^2) =
\int_{-i\infty}^{i\infty} \frac{d\lambda}{2 \pi i}
\;
e^{\lambda w^2} \; G_L(\lambda,t,t_w),
\end{equation}
we can extract its scaling properties from
the ones of $G_L(\lambda)$ in equation~(\ref{eq:GDttw}). Using
$y=\lambda w^2_\infty$ and
$2|c_n(0)|^2/w^2_\infty=
6w^2_0/(\pi^2n^2 w^2_\infty)=6/(\pi^2n^2) \,  s^2_0$
we find
\begin{eqnarray}
&&
w_\infty^2 \, P_L(w^2)= \Phi\left(
\frac{w^2}{w^2_\infty};\frac{\Delta t}{t_L},\frac{t_w}{t_L},s^2_{0} \right),
\label{eq:P1}
\\
&&
\Phi\left(x;\frac{\Delta t}{t_L},\frac{t_w}{t_L},s^2_{0}\right) =
\int_{-i\infty}^{i\infty} \frac{dy}{2 \pi i}e^{y\, x}
\prod_{n=1}^\infty \frac{ e^{-\frac{y\, s^2_{0}
b_n\left(\frac{\Delta t}{t_L},\frac{t_w}{t_L}\right)}{1+y\,
a_n\left(\frac{\Delta t}{t_L},\frac{t_w}{t_L}\right)}} }{1+y\,
a_n\left(\frac{\Delta t}{t_L},\frac{t_w}{t_L}\right)}
\label{eq:P11}
\end{eqnarray}
with the coefficients $a_n$ and $b_n$ defined in equations~(\ref{eq:an})
and (\ref{eq:bn}).
For a flat initial condition, $T_0=0$ and
$s_0^2=w_0^2/w_\infty^2=T_0/T=0$.
With a very similar calculation to the one
explained in~\cite{Zoltan} for the stationary case, we
rewrite the function $\Phi$ as
\begin{equation}
\label{eq:PhiT0}
\Phi\left(x;\frac{\Delta
t}{t_L},\frac{t_w}{t_L},0 \right)= \sum_{n=1}^\infty
\frac{e^{-x/a_n}}{a_n}
\prod_{m=1,m \neq n}^\infty
\frac{a_n}{a_n-a_m},
\end{equation}
which is essentially the same result in~\cite{Zoltan}
but with two-times dependent coefficients $a_n(\Delta
t/t_L,t_w/t_L)$.

By using now a different independent variable, $x'=w^2/\langle
w^2\rangle$, one has
\begin{eqnarray}
&&
\langle w^2\rangle
\; P_L(w^2)
=
\Phi'\left( \frac{w^2}{\langle w^2\rangle
};\frac{\Delta
t}{t_L},\frac{t_w}{t_L},s^2_0 \right),
\label{eq:PhiprimaEW}
\\
&&
\Phi'\left( x';
\frac{\Delta t}{t_L},\frac{t_w}{t_L},s^2_0 \right) \int_{-i\infty}^{i\infty} \frac{dy'}{2 \pi i} \; e^{y' x'} \,
\prod_{n=1}^\infty
\frac{e^{-\frac{y' s^2_0
b'_n\left(\frac{\Delta t}{t_L},\frac{t_w}{t_L}\right)}
{1+y' a'_n\left(\frac{\Delta t}{t_L},\frac{t_w}{t_L}\right)}}}
{1+y' a'_n\left(\frac{\Delta t}{t_L},\frac{t_w}{t_L}\right)}
\end{eqnarray}
and
\begin{eqnarray}
\label{eq:defanprim}
&&
a'_n= \frac{a_n}{\sum_{n=1}^\infty a_n + s^2_0 \sum_{n=1}^\infty b_n
},
\qquad
b'_n= \frac{b_n}{\sum_{n=1}^\infty a_n + s^2_0
\sum_{n=1}^\infty b_n }.
\end{eqnarray}
Equations~(\ref{eq:P1}) and (\ref{eq:P11}) are the generalization of equation~(15)
in~\cite{Zoltan} that takes into account the aging regime.
Equation~(\ref{eq:PhiprimaEW}) is a rewriting of the latter using the
more convenient normalized variable $w^2/\langle w^2\rangle$.  The
parameters are, in both cases, $\Delta t/t_L$, $t_w/t_L$ and $s_0^2$.
In the {\it growth and aging} regime in which $\Delta t/t_w$
is finite and the two parameters are very small, {\it i.e.} $\Delta t/t_L,
t_w/t_L \ll 1$, one formally has
\begin{eqnarray}
\;\; \frac{\Delta t}{t_w} = \epsilon, \qquad
\widetilde{w}^2 = {\cal G}(\epsilon), \qquad
\langle w^2\rangle = t_w^{1/2} \; {\cal G}(\epsilon).
\end{eqnarray}
We can now easily exchange
$\Delta t/t_L$ and $t_w/t_L$ by the more convenient set
$\widetilde{w}^2$ and $\langle
w^2\rangle/w_\infty^2$.
First, we exchange $\Delta t/t_L$ and $t_w/t_L$
by $\Delta t/t_w$ and $t_w/t_L$. Second,
on the one hand $\Delta t/t_w$ is an exclusive
function of $\widetilde{w}^2$. On the other hand, using
the results in Sect.~\ref{sec:scalrel} one can show that
\begin{eqnarray}
\frac{t_w}{t_L}
&=&
\left(
\frac{\langle w^2\rangle}{\widetilde{w}^2}
\frac{c_0(T,T_0)}{w_\infty^2}
\right)^2.
\end{eqnarray}
The factor $\widetilde{w}^2$ enters the last equation,
but we can ignore it by redefining the scaling function. We used the
EW exponents but this relation can be easily rewritten for
generic $\beta,\, z$ and $\zeta$. We thus have
\begin{equation}
\langle w^2\rangle P_L(w^2) = \Phi''\left(\frac{w^2}{\langle w^2\rangle};
\frac{\langle w^2\rangle}{w_\infty^2},
\widetilde{w}^2, T, T_0\right)
\end{equation}
as proposed in \cite{Bustingorryetal} for the generic disordered case.

\begin{figure}[!tbp]
\centerline{
\includegraphics[angle=-0,width=6cm,clip=true]{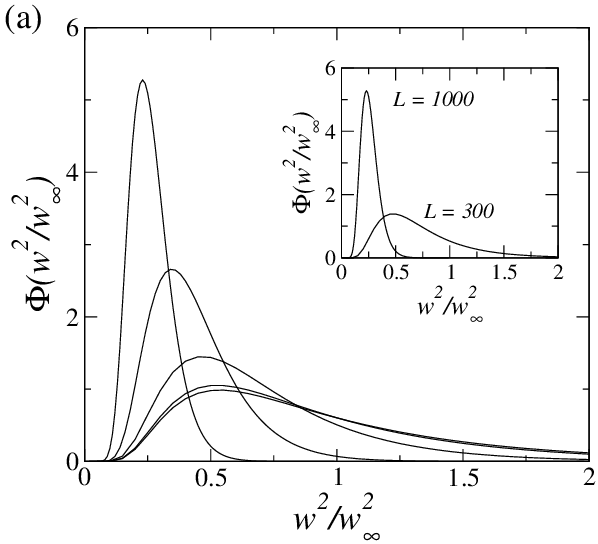}
\hspace{0.2in}
\includegraphics[angle=-0,width=6cm,clip=true]{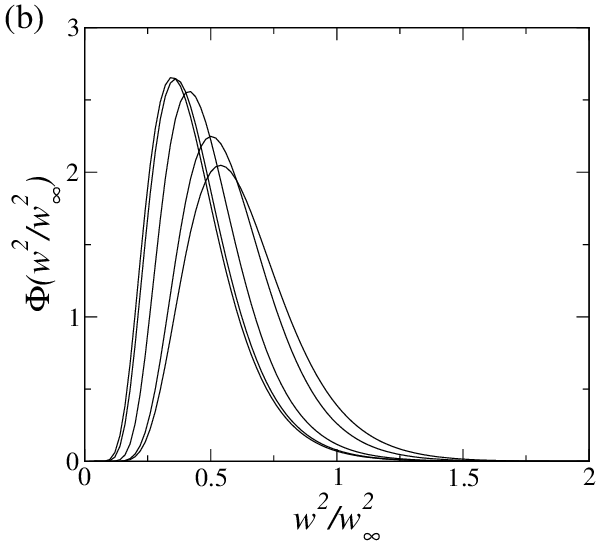}
}
\caption{\label{f:Phi_EW_ttw_a} Scaling function $\Phi(
w^2/w^2_\infty,\Delta t/t_L,t_w/t_L,0 )$ evaluated with $M_1=20$ and
$M_2=40$. (a) $L=1000$, $t_w=0$, and $\Delta
t=10^3,\,3\,10^3,\,10^4,\,3\,10^4$, and $10^5$ (from left to
right). The inset shows a comparison between $\Phi$ for different
system sizes, $L = 1000$ and $L = 300$ as indicated, at the same
waiting-time $t_w = 10^3$ (b) $L=1000$, $\Delta t = 3\,10^3$ and
$t_w=1,\,10,\,10^2,\,10^3$, and $10^4$ from left to right.}
\end{figure}

Let us list and illustrate in some plots different trends in the
scaling function $\Phi$ evaluated for the flat initial condition.
In the numerical evaluations we approximate
the infinite sums and products, as in~(\ref{eq:PhiT0}), by finite sums and products with
different cut-off values, $M_1$ and $M_2$ respectively.  The main
panel in Fig.~\ref{f:Phi_EW_ttw_a}~(a) shows the time-delay evolution
of the scaling function $\Phi$ for $t_w=0$ and fixed system size. This
corresponds essentially to the results obtained in \cite{Zoltan}, and
shows how the pdf is broader for increasing $\Delta t$ until the
saturation regime is reached. The inset shows the system size
dependence, indicating that at fixed $t_w$ the function $\Phi$, tends
to a delta-function in the infinite size limit.  In
Fig.~\ref{f:Phi_EW_ttw_a}~(b) one observes how the pdf is spread at
fixed $L$ and $\Delta t$ while increasing the waiting-time. These
results correspond to a flat initial condition $T_0=0$.

\begin{figure}[!tbp]
\centerline{
\includegraphics[angle=-0,width=6cm,clip=true]{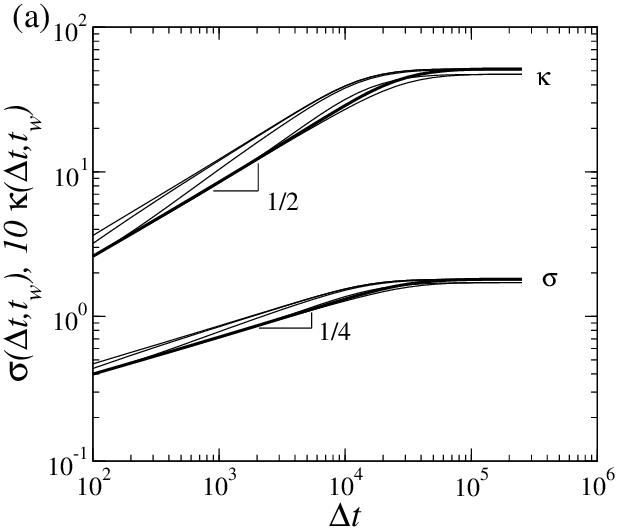}
\hspace{0.2in}
\includegraphics[angle=-0,width=6cm,clip=true]{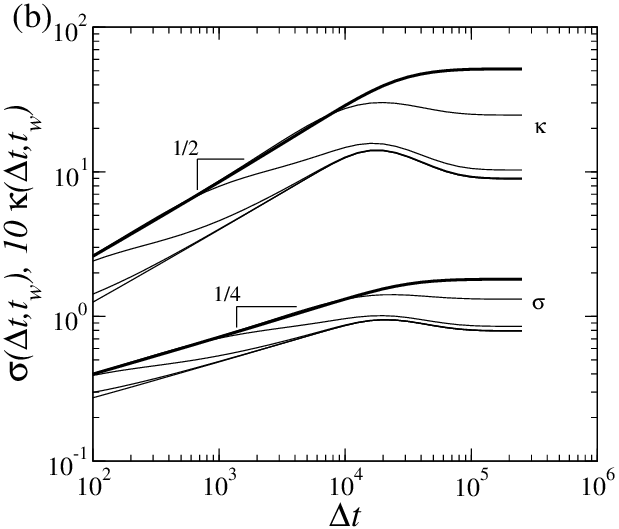}
}
\caption{\label{f:SK_EW_ttw_a} Two-times skewness $\sigma(\Delta
t,t_w)$ and kurtosis $\kappa(\Delta t,t_w)$ for the distribution
function $P_L(w^2)$ (the kurtosis is rescaled by a factor 10 for
clarity).  The sums are truncated with $M_3=100$. $T=1$, $L=1000$, and
$t_w=1,\,10,\,10^2,\,10^3,\,10^4$, and $10^5$ as indicated. Different
initial conditions correspond to (a) $T_0=0$ and (b) $T_0=5$.  The
thick lines correspond to the $t_w \rightarrow \infty$ limit.}
\end{figure}

\begin{figure}[!tbp]
\centerline{
\includegraphics[angle=-0,width=8cm,clip=true]{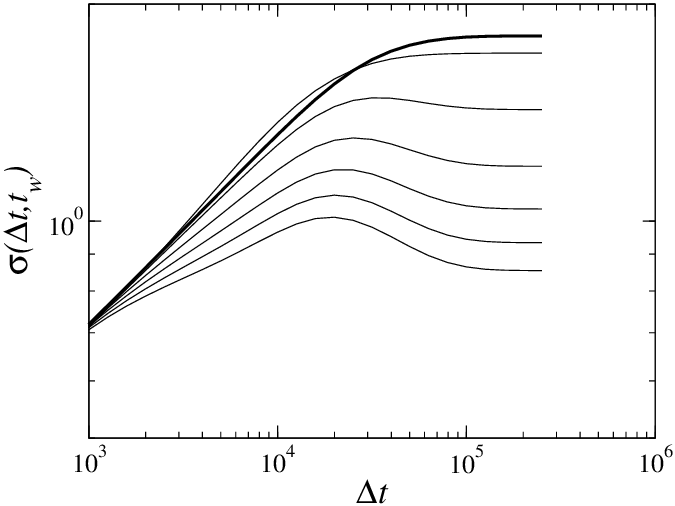}
}
\caption{\label{f:SK_EW_ttw_b} Two-times skewness $\sigma(\Delta
t,t_w)$ for the distribution function $P_L(w^2)$.  The sums are
truncated with $M_3=100$. $T=1$, $L=1000$, and $T_0=
0,\,1,\,2,\,3,\,4$, and $5$ from top to bottom. The waiting-time
is $t_w=10^3$. For comparison, the $t_w=10^5$ curve, which is
the same for all initial temperatures is also included as a thick line.}
\end{figure}

Since it is hard to compute the scaling function $\Phi$
for a $T_0>0$ initial condition we just present
the skewness and kurtosis,
\begin{equation}
\sigma = \frac{\mu_3}{\mu_2^{3/2}},
\qquad
\kappa = \frac{\mu_4}{\mu_2^2}-3,
\end{equation}
respectively, with the centered moments defined as $ \mu_2 = \langle
w^4 \rangle - \langle w^2 \rangle^2$, $\mu_3 = \langle w^6 \rangle - 3
\langle w^4 \rangle \langle w^2 \rangle + 2 \langle w^2 \rangle^3$,
and $ \mu_4 = \langle w^8 \rangle - 4 \langle w^6 \rangle \langle w^2
\rangle + 6 \langle w^4 \rangle \langle w^2 \rangle^2 - 3 \langle w^2
\rangle^4$. The moments  of the roughness pdf
are given by
$\langle (w^2)^m \rangle = (-1)^m \left. \partial_{\lambda^m}
G_L(\lambda)\right|_{\lambda = 0}$;
then after some algebra one finds
\begin{eqnarray}
\sigma = \frac{\sum_{n=1}^\infty \left[ 2 a_n^3 + 6 s_0^2
a_n^2 b_n \right]}{\left[ \sum_{n=1}^\infty \left( a_n^2 + 2 s_0^2 a_n
b_n \right) \right]^{3/2}},
\qquad
\kappa = \frac{\sum_{n=1}^\infty \left[ 6 a_n^4 + 24 s_0^2 a_n^3 b_n  \right]}
{\left[ \sum_{n=1}^\infty \left( a_n^2 + 2 s_0^2 a_n b_n \right) \right]^2}.
\end{eqnarray}
Figure~\ref{f:SK_EW_ttw_a} displays the time-delay evolution of the
skewness and kurtosis for the cases $T>T_0=0$
[Fig.~\ref{f:SK_EW_ttw_a}~(a)] and $T<T_0$
[Fig.~\ref{f:SK_EW_ttw_a}~(b)]. One can observe that the pdfs are
broader and more asymmetric until saturation. Generally, both $\sigma$
and $\kappa$ age with a similar $\Delta t$ and $t_w$ dependencies
as the correlation
length $l$ or the averaged roughness $\langle w^2\rangle$.
The peculiarities are that the
asymptotes corresponding to the infinite size limits grow as $\Delta
t^{1/4}$ for the skewness and $\Delta t^{1/2}$ for the kurtosis. The
approach to saturation is non-monotonic, showing a bump around $\Delta
t \approx t_x(t_w)$. Finally, Fig.~\ref{f:SK_EW_ttw_b} shows the
$T_0$-dependence of the skewness for $t_w=10^3\ll t_L$ (the kurtosis
behaves in a similar way).

\subsection{Center-of-mass displacement distribution}

In Sect.~\ref{sec:MSD} we showed that the line's
center-of-mass undergoes Brownian motion with mean
$\overline{x}(0)$
and variance $\langle D\rangle (\Delta t)=2T/(\gamma L)\Delta t$.
Thus, the center of mass position is Gaussian
distributed
and, after a change of variables,
the distribution of the mean-squared-displacement of the
center-of-mass is
\begin{equation}
\label{eq:PdeD}
P_L(D)=\frac{e^{-D/(2 \langle D \rangle)}}{\sqrt{2 \pi \langle D \rangle
D}}= \sqrt{\frac{\gamma L}{4 \pi T \Delta t D}} \exp \left(
-\frac{\gamma L}{4 T \Delta t} D \right)
\end{equation}
which can also be written in the scaled form
\begin{equation}
\langle D\rangle \, P_L(D)= \Omega \left( \frac{D}{\langle D \rangle}\right),
\qquad
\mbox{with}
\qquad
\Omega\left(x\right) = \frac{e^{-\frac{x}{2}}}{\sqrt{2 \pi x}}.
\end{equation}

\subsection{Mean-squared displacement distribution}

The mean-squared-displacement
satisfies $B(\Delta t,t_w)=w^2(\Delta
t,t_w)+D(\Delta t)$. $w^2$ and $D$ are independent variables
(note that the roughness is independent of the zero mode), then the
probability function for $B$ can be formally obtained from the inverse
Laplace transform of the product of Laplace transforms,
\begin{equation}
P_L(B)=\int_{-\infty}^{\infty}\frac{d \lambda}{2 \pi i}e^{\lambda B}
K_L(\lambda,\Delta t,t_w),
\end{equation}
where $K_L=G_L\,J_L$ is the Laplace transform of $P_L(B)$,
$G_L(\lambda,\Delta t,t_w)$ is given by equation~(\ref{eq:GDttw}), and
$J_L(\lambda,\Delta t)$ is the Laplace transform of $P_L(D)$,
\begin{equation}
J_L(\lambda,\Delta t)= \frac{1}{\sqrt{2 \langle D \rangle \lambda +1}}
= \sqrt{\frac{\gamma L}{4 T \Delta t \lambda + \gamma L}}.
\end{equation}
Then one finds that
\begin{equation}
\fl
K_L(\lambda,\Delta t,t_w) =
\frac{1}{\sqrt{2 \left[ \langle B \rangle - \langle w^2 \rangle \right] \lambda
+1}}\;
\prod_{n=1}^\infty
\frac{e^{-\frac{2 \lambda |c_n(0)|^2 \left( 1- e^{-\nu k_n^2
\Delta t} \right)^2 e^{-2 \nu k_n^2 t_w}}
{1+\lambda w_\infty^2 a_n(\Delta t,t_w)}}}
{1+\lambda w_\infty^2 a_n(\Delta t,t_w)}.
\end{equation}
We now define $y''=\lambda \langle B
\rangle$, $\lambda w_\infty^2 a_n(\Delta t,t_w) = y''
a''_n(\Delta t,t_w)$, and
\begin{eqnarray}
&&
a''_n\left(\frac{\Delta t}{t_L},\frac{t_w}{t_L}\right)
=
\frac{a_n}{\langle B \rangle / w_\infty^2},
\qquad
b''_n \left(\frac{\Delta t}{t_L},
\frac{t_w}{t_L}\right)
=\frac{b_n}{\langle B \rangle /w_\infty^2},
\end{eqnarray}
and we obtain
\begin{equation}
\langle B \rangle P_L(B)=
\Psi \left(\frac{B}{\langle B
\rangle};\frac{\Delta t}{t_L},\frac{t_w}{t_L},s_0^2 \right).
\end{equation}
The scaling function is given by
\begin{equation}
\fl
\Psi\left(x''; \frac{\Delta
t}{t_L},\frac{t_w}{t_L},s_0^2 \right)=
\int_{-i \infty}^{+i \infty}\frac{d y''}{2 \pi i}e^{y'' x''}
\;
\prod_{n=1}^\infty \frac{1}{\sqrt{1 + 2 c''_n y''}}
\frac{e^{ -\frac{y'' s_0^2 b''_n(\Delta t,t_w)}
{1+y'' a''_n(\Delta t,t_w)}}}{1+y''a''_n(\Delta t,t_w)}
,
\end{equation}
with
\begin{equation}
c''_n\left(\frac{\Delta t}{t_L},\frac{t_w}{t_L}\right)
=
\frac{\langle D \rangle}{\langle B \rangle}.
\end{equation}
The pdf of the mean-squared-displacement $B$
can also be written in a scaling form similar to the one found for the
roughness.

\subsection{The incoherent scattering function}

In order to obtain the pdf of the incoherent scattering
function we need to use the moment-expansion:
\begin{equation}
p(z) = \int \frac{du}{2\pi} \; e^{i z u} \, p(u)
\end{equation}
with
\begin{equation}
p(u) = \sum_{p=0}^\infty \frac{u^p}{p!} \;
\left. \frac{\partial^p p(u)}{\partial u^p} \right|_{u=0}
=
\sum_{p=0}^\infty \frac{(-iu)^p}{p!} \;
\langle \, z^p \, \rangle
\; .
\end{equation}
This expression assumes that the series converges and the moments
exist.
The moments of $C_q$ are
\begin{equation}
\langle \, C_q^p\, \rangle
=
L^{-p} \int dz_1 \dots dz_p \;
e^{-\frac{q^2}{2} \sum_{r,r'=1}^p \langle H \rangle (z_r-z_{r'};t,t_w)},
\end{equation}
where the function
\begin{eqnarray}
\label{eq:HHcorr}
& \langle H \rangle(z;t,t_w)
= \frac{2}{L}\sum_{n=1}^\infty \langle S_n \rangle (\Delta t,t_w) e^{ik_n z}
\nonumber \\
&  \quad =L^{-1} \int_0^L dz' \;
\langle
\left[
\delta x(z',t) - \delta x(z',t_w)
\right]
\left[
\delta x(z'-z,t) - \delta x(z'-z,t_w)
\right]
\rangle
\end{eqnarray}
is the two-times generalization of the height-height correlation function~\cite{loschinos}.

Using the fact that $\langle H \rangle (0;t,t_w)=\langle w^2 \rangle
(\Delta t, t_w)$ and equation~(\ref{eq:CqGauss}), one can show that
$\langle \, C_q^p\, \rangle = \langle C_q \rangle^p I_p$, where the
$q$-dependent function $I_p(t,t_w)$ is given by
\begin{equation}
I_p(\Delta t,t_w,q^2L)=L^{-p}\int dz_1 \dots dz_p \;
e^{-\frac{q^2}{2} \sum_{r,r'=1;r \neq r'}^p \langle H \rangle(z_r-z_{r'};\Delta t,t_w)}.
\end{equation}
From the moment-expansion, using $u'=u\langle C_q \rangle$, the
fact that $T q^2 L$ is a function of $\langle C_q^\infty \rangle$ and
calling $T_0 q^2 L=\langle C_q^0\rangle$, one can write the pdf of the
incoherent scattering function in the scaled form
\begin{equation}
\langle C_q \rangle
P\left( C_q \right)
= \Theta\left(\frac{C_q}{\langle C_q \rangle};
\frac{\Delta t}{t_L}, \frac{t_w}{t_L}, C_q^\infty, C_q^0 \right),
\end{equation}
with
\begin{equation}
\Theta\left(x;\frac{\Delta t}{t_L}, \frac{t_w}{t_L}, C_q^\infty, C_q^0 \right) =
\sum_{p=0}^\infty \frac{(-i)^p}{p!} \int \frac{du'}{2 \pi}e^{i u' x} \; u'^p \;
I_p\left( \frac{\Delta t}{t_L}, \frac{t_w}{t_L},q^2 L \right).
\end{equation}
Although in the last expression the functional form of $\Theta(x)$ is
not evident, the scaling properties are clear.

\subsection{Fluctuations of the response functions}

In quadratic models as the EW elastic line the response
functions do not fluctuate. This can be easily shown as
follows. Take the roughness integrated linear response
(\ref{eq:resp-w2}) without the thermal average. Replacing
$c^h_n(t)$ and $c_n(t)$ by their functional form, and using the
fact that the two copies evolve with the same thermal noise,
one has
\begin{equation}
\chi^{w^2}(t,t_w) = 2L \sum_{n=1}^\infty s_n^2 \, e^{-\nu k_n^2 (t-t_w)}
\; .
\end{equation}
This result depends on the random fields $s_n$ but it is independent
of the thermal noise. For {\it fixed} random fields this quantity does
not fluctuate and the pdf of $\chi^{w^2}$ is a delta function.
Similarly, one can prove that the displacement and center-of-mass
responses are delta-distributed. The same `trivial' result was found
for the ferromagnetic coarsening in the O($N$) model with
$N\to\infty$~\cite{Chcuyo}.

\section{Summary and conclusions}
\label{sec:summary}

We studied the averaged and fluctuating dynamics of the
Edwards-Wilkinson elastic line in one transverse dimension.

Firstly, we analyzed the evolution of correlation functions in terms
of the different time scales involved in the problem: the total time,
$t$, the waiting time, $t_w$, and the saturation time $t_L$. In 
particular, for the two-times roughness we found 
$\langle \, w^2 \, \rangle \sim {\cal F}_{w^2} \left[
\Delta t/t_w,t_w/t_L, TL, T_0L \right]\sim {\cal F}_{w^2}
\left[t/t_w,t_w/t_L, TL, T_0L \right]$ (the scaling
function in the third member is not identical to the one in the second
member but we use the same name to simplify the notation).  
As mentioned earlier, the problem can also be analyzed in terms of
associated length scales obtained from the growing correlation length
$\ell(t)=4\pi^2\nu \,t^{1/2}$.  Then, the relevant length scales are
$\ell(t)$, $\ell(t_w)$, and $\ell(t_L)\sim L$.  This is reflected for
instance in the scaling form of the structure factor in the asymptotic
time-delay limit (\ref{eq:scaling-S}), using $k_w \sim
\ell(t_w)^{-1}$. The scaling form of the two-times averaged roughness
that describes the 
aging, saturation and equilibrium regimes can be written as  
\begin{equation}
\langle \, w^2 \, \rangle(\Delta t,t_w)
\sim
{\cal F}_{w^2} 
\left[ \frac{\ell(t)}{\ell(t_w)},\frac{\ell(t_w)}{\ell(t_L)}, TL, T_0L \right]
\label{eq:fitting1}
\end{equation}
with $\ell(t) =4\pi^2\nu t^{1/2}$
and $t_L=L^2/(4\pi^2\nu)$
in the $1+1$ EW case.  This form extends the proposal in
(\ref{eq:fitting0}) to include another scaling variable and thus
describe the dynamics of finite lines.  It then generalizes the
celebrated Family-Vicsek scaling~\cite{Favi} to include the
preasymptotic non-equilibrium regime.  The aging regime corresponds to
$\ell(t_w)\ll \ell(t_L)$ and the function ${\cal
F}_{w^2}[\ell(t)/\ell(t_w),0,T_0L,TL] \sim \ell^\zeta {\cal F}[\ell(t)/\ell(t_w)]$,
leading to (\ref{eq:fitting0})
with all the temperature dependent 
asymptotic properties  already
detailed in the central part of the manuscript. 
 The saturation regime
is reached by taking $t\gg t_L$ at fixed $t_w$; this means 
$\ell(t_w)/\ell(t) \ll 1$ and $\ell(t)/\ell(t_L) \gg 1$.  
Finally, the usual stationary equilibrium regime corresponds to $t_w
\gg t_L$, and for a power-law growth one recovers the Family-Vicsek
scaling $\langle w^2 \rangle \sim L^\zeta {\cal F}_{w^2}[\ell(\Delta
t)/L]$.

We also showed that the two-times dependent correlation length defined
through the dynamics of the structure factor, satisfies a similar
scaling law
\begin{equation}
l(t,t_w) \sim {\cal F}_l\left[
\frac{\ell(t)}{\ell(t_w)},\frac{\ell(t_w)}{\ell(t_L)},\frac{T}{T_0} \right].
\label{eq:fitting2}
\end{equation}
Note that
although this expression gives the full aging behaviour of $l(t,t_w)$,
it can be completely rationalized using the simple length scale
$\ell(t)$. For instance, in the aging regime, one finds that regions
with $\ell(t) \sim \ell(t_w)$ are equilibrated at the
working temperature, while regions with $\ell(t) > \ell(t_w)$
are still not at equilibrium.

Interestingly enough, we demonstrated that ordering or disordering
non-equilibrium dynamics following a quench from higher temperature or
a heating process from a lower temperature are characterized by a
higher or lower effective temperature than the one of the bath. This
result is consistent with the intuitive interpretation of the
effective temperature with higher (lower) values associated to more
(less) disordered configurations than the equilibrium ones at the
working temperature. A similar dependence on the initial condition was
derived by Berthier {\it et al} in the $2d$ XY model~\cite{Behose}.

The two-times length scale $l$ also shows the latter property.  In the
aging regime, for fixed $t_w$ the length grows with $\Delta t$ for all
$T\neq T_0$ while for fixed $\Delta t$ it grows with $t_w$ when
$T_0>T$ and it decreases with $t_w$ when $T_0<T$. The former behaviour
is similar to what is observed in conventional glassy systems such as
the $3d$ Edwards-Anderson spin glass~\cite{Ludovic} and models of
particles in interaction~\cite{Parisi}. The heating procedure was not
studied in such cases.

The wave-vector dependent correlation $\langle C_q\rangle $ that plays
the role of the incoherent scattering function in studies of glassy
systems is particularly interesting.  We showed that, although
$\langle C_q\rangle $ is simply related to the roughness, the
characteristic multiplicative scaling is not easily detected in
$\langle C_q\rangle $. This might be the case in other systems, such
as colloidal glasses where it was recently shown that diffusive
correlations clearly display multiplicative aging
scaling~\cite{Makse}, while this was not previously reckoned in the
study of the incoherent scattering function~\cite{laponite}.

One can also observe that the aging behaviour of $\langle C_q\rangle $
resembles strongly the experimental results in
laponite~\cite{laponite}.  In particular it was found in this system
that the incoherent scattering function displays a waiting-time
dependent plateau at long time-delay. This suggests that a similar
equilibration mechanism might be at work in the relaxation dynamics of
the rather complex laponite samples, where some competing length scale
is confining the particle fluctuations, thus leading to saturation of
the incoherent scatering function. The waiting-time dependent
saturation is in line with the fact that effective temperatures,
measured through the FDT, should become the bath temperature at fixed time-delay and
sufficiently long waiting times where saturation is found.  The
measurements in~\cite{Abou} are such that the effective temperature
does indeed tend to the bath temperature at fixed working frequency --
equivalently time-delay -- when $t_w$ is large enough, although this
result remains controversial~\cite{controAbou}.

With this analytic study we showed that the qualitative aging dynamics
of the vortex glass~\cite{us,schehr05} as well as the elastic line in
a quenched disordered environment~\cite{Yosh98,Bustingorryetal} is
mainly due to the non-equilibrium relaxation of the pure elastic
lines. The effect of quenched disorder and line-line interactions is
to change the details of the scaling, more precisely the temperature
and time-dependence in $\ell(t)$, but not the qualitative features.
Along this line, the study of the aging dynamics of the pure KPZ
equation~\cite{BustKPZ} will allow one to better rationalize the
results in~\cite{Spaniards}, where the aging dynamics of this equation
with a disordered potential -- and driving force -- was analyzed.
Furthermore, the results obtained here could be strongly related to the
non-equilibrium relaxation dynamics of confined polymers~\cite{claudio}.

We presented the first analytic calculation of finite-size
fluctuations during an out-of-equilibrium relaxation. This study
complements the analysis in \cite{static-inter} and \cite{Zoltan} for
the width fluctuations at saturation and growth and in
\cite{Satya,Greg} for other quantities such as the maximum height
displacement -- indeed, it is also simple to include the $t_w$-dependence 
in this calculation.  Our results make explicit the dependence on the
waiting-time and display the crossover to equilibrium. They
constitute a benchmark for R\'acz proposal to classify interface dynamics
into universality classes~\cite{Racz}, now extending it to the
non-equilibrium relaxation.

As regards the time-reparametrization invariance scenario for glassy
dynamics we do not expect it to hold, without modification, in models
with multiplicative aging scaling. Following the steps sketched in
\cite{Chamon-Cugliandolo} to study the asymptotic averaged dynamics of
the EW line (or massless scalar field) and its corresponding dynamic
action, one soon realizes that the multiplicative $t_w^{1/2}$ factor
has to be extracted from the asymptotic analysis to search for
time-reparametrization invariance. This is similar to what was shown
in \cite{Chcuyo} for the O($N$) model in the large $N$ limit.  One
should also notice that the dynamics of the EW line depends on the
dimensionality of the transverse space. For instance, one can show
that for infinite systems diffusion disappears and the aging regime
persists at infinite waiting-times ({\it i.e.}  the scaling becomes
additive) in two transverse dimensions~\cite{Cukupa}. One has then the
interesting possibility of testing the time-reparametrization
invariance scenario in the `critical' EW equation with two transverse
dimensions.  The detailed analysis of the dynamic symmetries of the
generic EW line goes beyond the scope of this article.

We conclude with a note on the relevance of our results for coarsening
phenomena. The domain walls between equilibrated regions during domain
growth  are usually described as elastic objects. Recently,
the distribution of domain sizes and perimeter lengths in
two-dimensional Ising ferromagnetic growth was shown to be
unexpectedly non-trivial~\cite{Alberto}.  The wide distribution of
domain sizes and domain wall lengths combined with the highly
non-trivial fluctuating dynamics of finite-length elastic lines
derived here suggest that characterising the fluctuations of standard
two-times observables in domain-growth phenomena can be 
a quite complicated problem.

\vspace{0.2cm} \ack We thank the Universidad Nacional de Mar del
Plata, Argentina, for hospitality during the preparation of this work
and C. Chamon, D. Dom\'{\i}nguez, T. Giamarchi, G. Schehr and
H. Yoshino for very useful discussions.  LFC acknowledges financial
support from Secyt-ECOS P. A01E01 and PICS 3172, SB from the Swiss
National Science Foundation under MaNEP and Division II, and JLI from 
CONCIET PIP05-5648 and ANPCYT PICT04-20075. LFC is a
member of Institut Universitaire de France.

\end{document}